\documentclass{article}

\usepackage{arxiv}

\usepackage{xcolor,soul,framed} %,caption
\colorlet{shadecolor}{yellow}
\usepackage[pdftex]{graphicx}
\graphicspath{{../pdf/}{../jpeg/}}
\DeclareGraphicsExtensions{.pdf,.jpeg,.png}
\usepackage{multirow}
\usepackage{amsmath}
\usepackage{setspace}
\usepackage{amsfonts}
 \usepackage{comment}
\usepackage{array}
\usepackage{mathtools}
\usepackage{mdwmath}
\usepackage{cite}
\usepackage{mdwtab}
\usepackage{eqparbox}
\usepackage{url}
\usepackage{siunitx}
\DeclareMathOperator*{\Motimes}{\text{\raisebox{0.25ex}{\scalebox{0.75}{$\bigotimes$}}}}

\newcommand{\hb}{\mathbf}

\newcommand{\be}{\begin{equation}}
\newcommand{\ee}{\end{equation}}
\newcommand{\bi}{\begin{itemize}}
\newcommand{\ei}{\end{itemize}}
\newcommand{\bn}{\begin{enumerate}}
\newcommand{\en}{\end{enumerate}}
\newcommand{\bea}{\begin{eqnarray}}
\newcommand{\eea}{\end{eqnarray}}
\newcommand{\bef}{\begin{figure}}
\newcommand{\ef}{\end{figure}}

\usepackage{algorithm}
\usepackage{algpseudocode}

\usepackage{amssymb}

\usepackage{subfig}
\usepackage{graphicx}

\title{Efficient Algorithms for Convolutional Inverse Problems in Multidimensional Imaging}

\author{
  Didem~Dogan\thanks{D.~Dogan was with the Department of Electrical and Electronics Engineering, METU, Cankaya, Ankara, 06800 Turkey.} \\
  Circuits and Systems Group\\
  Delft University of Technology\\
  Delft, 2628, Netherlands\\
  \texttt{d.doganbaskaya@tudelft.nl}  \\
   \And
 Figen~S.~Oktem \\
  Department of Electrical and Electronics Engineering\\
  Middle East Technical University (METU)\\
  Ankara, 06800, Turkey \\
  \texttt{figeno@metu.edu.tr} \\
}

\begin{document}
\maketitle

\begin{abstract}
% Computational imaging is the process of indirectly forming images from measurements using image reconstruction algorithms that solve inverse problems. 
Multidimensional imaging, capturing image data in more than two dimensions, %(unlike conventional photography), 
has been an emerging field with diverse applications.
%with application in diverse fields such as %widespread applications in 
%physics, chemistry, biology, medicine, and remote sensing.
Due to the %intrinsic 
limitation of two-dimensional detectors
in obtaining %inherently three-dimensional (3D) data,
the %multidimensional
high-dimensional image data, 
computational
imaging approaches 
%emerges as an effective approach by passing on some of the
have been developed to pass on some of the
burden to a %computational system.
reconstruction algorithm.
In various %many %inverse 
image reconstruction problems in %computational 
multidimensional imaging, % (such as spectral and depth imaging), 
the measurements are in the form of superimposed convolutions. % - related to the unknown image. 
%of multidimensional image data
In this paper, we introduce a general framework for the solution of these problems, called here \textit{convolutional inverse problems}, and develop fast image reconstruction algorithms %that exploit sparse models in analysis and synthesis forms. % for regularization. 
with analysis and synthesis 
%sparsity models. These priors 
priors. These 
include sparsifying transforms, as well as %data-adaptive
convolutional or patch-based dictionaries that %are patch-based or convolutional. 
can adapt to correlations in different dimensions.
%can adapt the corrrelations in the unknown image data that are in various dimensions
The resulting %regularized inverse 
optimization problems are solved via alternating direction method of multipliers %(ADMM)
with closed-form, efficient, and parallelizable update steps. 
%Finally, the developed algorithms with different priors are illustrated for three-dimensional image reconstruction problems in computational spectral imaging, and their performance is comparatively evaluated for cases with or without correlation along the third dimension. 
To illustrate their utility and versatility, the developed algorithms are applied to three-dimensional image reconstruction problems in computational spectral imaging for cases with or without correlation along the third dimension.
%This work opens up new possibilities for fast iterative reconstruction in large-scale imaging problems.
As the advent of multidimensional imaging modalities expands to perform sophisticated tasks, these algorithms are essential for fast iterative reconstruction in various large-scale problems.
\end{abstract}

% keywords can be removed

\section{Introduction}

%\begin{itemize}
%    \item  Multidimensional Imaging:
%    \item  Convolutional Inverse Problems in Multidimensional Imaging:
%    \item  Conventional sparsity based methods: TV, DCT, analysis, synthesis methods
%    \item  Development of learning based methods.
%    \item  Convolutional Sparse Representations without d update, with d update
%    \item  Contributions of this work
%\end{itemize}
%

Multidimensional imaging, that is,
capturing %electromagnetic fields
image data in more than two dimensions,
% capturing the multidimensional information of an object
%(unlike conventional photography), 
has been a
prominent %emerging 
field with ubiquitous %widespread
applications in
%physics, chemistry, biology, medicine, astronomy,
%and remote sensing
the physical and life sciences~\cite{gao2016,oktem2018_chapter}. The %measured 
multidimensional image
data, including the spatial, spectral,
and temporal distributions of light (or %other 
an electromagnetic field), provide
unprecedented information about the chemical,
physical, and biological properties of targeted
%objects
scenes~\cite{gao2016,oktem2018_chapter,willett2014,schlemper2017,antipa2018}. %gao2014,
%Gao2014 : Nature paper: single-shot ultrafast photography
%Cao2016: Computational Snapshot Multispectral Cameras
%schlemper2017: A deep cascade of convolutional neural networks for dynamic MR image reconstruction 

While the objective of conventional photography is to measure only the
two-dimensional spatial distribution of light, the objective of multidimensional imaging is to form images of a radiating scene as a function of more than two variables. That is, the goal is to obtain a datacube of high dimensions, for example, in three spatial coordinates $(x, y, z)$, wavelength $(\lambda)$ and time $(t)$. However, obtaining this high-dimensional image data with inherently two-dimensional detectors poses intrinsic limitations on the spatio-spectral-temporal extent of these techniques.

Conventional techniques circumvent this limitation by sequential scanning of a series of two-dimensional measurements to form the high-dimensional image data. For example, in spectral imaging, the three-dimensional datacube $(x, y, \lambda)$ is typically obtained by either using a spectrometer with a long slit and scanning the scene
spatially, or by using an imager with a series of spectral filters and scanning the scene spectrally. 

%In the former case only a thin slice of the scene
%is observed at a time, whereas in the later case only one spectral band is observed at a time. 
%As a result, 
As a result, these scanning-based conventional methods
%sacrifice %light throughput, 
generally suffer from low signal-to-noise ratio (SNR), high acquisition time, and temporal artifacts for dynamic scenes.
%,  and also are best suited for imaging of scenes that
%remain stationary during the scanning process involved. For dynamic scenes, however, the conventional methods are subject to temporal artifacts.
Moreover, the attainable resolutions (such as temporal, spatial, and spectral) are inherently limited by the physical components involved.
%The conventional multidimensional imaging techniques also have inherent limitations on the attainable temporal, spatial, and spectral resolutions as imposed by their reliance on purely physical measurement systems.

To overcome these drawbacks, computational imaging approaches have been developed to pass on some of the burden to a reconstruction algorithm~\cite{oktem2018_chapter, schlemper2017, oktem2014, antipa2018,kar2019}. In these approaches, image data is reconstructed by combining information from multiplexed measurements with the additional prior (statistical or structural) knowledge about the unknown image.  

In many image reconstruction problems in multidimensional imaging, % (such as spectral and depth imaging), 
the measurements are in the form of superimposed convolutions. That is, the relationship between the measured (sensor) data and the unknown images can often be adequately characterized by %a single convolution operation or 
sum of multiple convolutions. In fact, for linear shift-variant systems whose response slowly varies across the field of view, %space
time, %instrument field of view
depth, or spectral dimensions,
%(That is, a linear shift-variant imaging system
%a linear 
the system operator can often be approximated by a linear combination of regular convolution operators~\cite{denis2015,sroubek2016}. 
%  as a WEIGHTED sum of convolutions 
%the image-formation model can be % well
%approximated by a linear and shift-invariant model). 
In this paper, we focus on the solution of this type of inverse problems, which are called here \emph{convolutional inverse problems}~\cite{dogan2019}.  
%These problems are named as convolutional inverse problems~\cite{dogan2019}. 
%Examples of such inverse problems are encountered in 
Such inverse problems are encountered in various computational imaging modalities such as
computational photography,
wide-field astronomical imaging, three-dimensional microscopy, spectral imaging, ultrafast imaging, radio interferometric imaging, magnetic resonance imaging, and ultrasound imaging~\cite{ dogan2019,
antipa2018, oktem2014, kar2019, camera2015, sroubek2016, denis2015, zhang2010,  badali2017, rau2011, stewart2011, kim2012,gol2016, florea2018, besson2019}.

In this paper, we %first 
introduce a unified framework for the solution of convolutional inverse problems by %defining
considering a general image-formation model. Based on alternating  direction  method  of multipliers (ADMM)~\cite{boyd2011}, we  
develop fast image reconstruction algorithms that can exploit sparse models in analysis or synthesis forms for the high-dimensional image data, as well as correlations in different dimensions. 
% for regularization.
%with analysis and synthesis sparsity models.
In the analysis case, multidimensional discrete derivative operators or sparsifying transforms can be utilized such as discrete cosine transform (DCT), wavelets, or their %high-dimensional 
Kronecker-product forms~\cite{elad2007, antipa2018, kar2019}. %duarte2012, 
% duarte2012: Kronecker Compressive Sensing
In the synthesis case, convolutional or patch-based dictionaries can be utilized, which can also be adapted to correlations in different dimensions~\cite{dogan2019,aharon2006, ravishankar2010, caballero2014, lorintiu2015, wohlberg2016, hu2017, cardona2018, nguyen2019, barajas2020,rey2020}. 
%%%REVISION: Yukaridaki referanslar azaltilabilir.
%If the image of interest is correlated in all directions, then high-dimensional transforms or dictionaries can be used in these models.
%For example, 
Based on the available prior knowledge about the image of interest, there may be correlations either 
all through the image data %in all directions 
%exploit the correlations among all the dimensions simultaneously
or only in certain dimensions.
The inverse problem is formulated for both cases and the resulting %regularized inverse 
optimization problems are solved via ADMM. The obtained reconstruction algorithms have closed-form, efficient, and parallelizable update steps.
To illustrate their utility and versatility, these %developed
algorithms %that allow different priors 
are %investigated
%evaluated 
%also 
applied to three-dimensional (3D) %image 
reconstruction problems  
%for images correlated along the third dimension
in computational spectral imaging, and their performance is numerically demonstrated for various cases with or without correlation along the third dimension.

ADMM-based reconstruction algorithms have been earlier developed for certain multidimensional imaging modalities with specific prior and observation models~\cite{antipa2018, kar2019}. 
%  chen2018 
% serrano2018, solano2019,
%solano2019:  %  "Tridimensional Convolutional Sparse Coding of Spectral Images," denoising. ADMM kullanılıyor, prior ise convolutional dictionary 
% chen2018:  "Quantitative differential phase contrast (DPC) microscopy with computational aberration correction,"  combinations of convolutions var ama tam istediğimiz stilde değil bence. ADMM kullanılıyor.
% serrano2018: "Convolutional sparse coding for high dynamic range imaging"
% burda sum of convolutions yok forward modelda tek bir convolution var (eğer bunu koyarsak K = 1 ve S = 1 için başka şeyler de koymamız gerekebilir) convolutional dictionary kullanmış, ADMMle çözmüş.  
% Buraya bizden baska Diffusercam makalesi girecek. Baska girmesi gereken makale var mi? Yani, "convolutional forward model+ADMM based reconstruction" iceren baska makale var mi literaturde? Su baglantida cikan makalelerin hepsine bu gozle dikkatlice bakmanda yarar var:
% https://scholar.google.com/scholar?hl=en&as_sdt=0%2C5&q=%22sum+of+convolutions%22+OR+%22superposition+of+convolutions%22+OR+%22combination+of+convolutions%22+reconstruction+ADMM&btnG=
% BU SORUNUN CEVABINI BANA LUTFEN EPOSTA ILE DE GONDER.
%%%%%%%%%%%%%%%%%%%%%%%%%%%
% FIGEN-TODO: Add here a comment about kronecker-based, tensor-based approaches...
%Another related field of research is ....
To the best of our knowledge, convolutional inverse problems have not been studied with this generality %before
% unified treatment of these problems
considering different %various 
correlation and prior models. 
%priors with different correlation forms (with various correlation forms and prior models) 
A multidimensional signal of interest may or may not be correlated in all directions,
%among all the dimensions
which accordingly determines the dimension %size
of the transforms/dictionaries 
%requires to use multidimensional transforms or dictionaries 
used in these models (for example, 2D or 3D) and the reconstruction approach.  
The versatile ADMM-based reconstruction algorithms developed in this paper are %efficient and 
powerful in that they can be applied to various imaging modalities involving convolutional inverse problems. 

This paper is organized as follows. In Section II, we introduce the convolutional %forward
measurement model. The convolutional inverse problem is then formulated in Section III using different priors. %sparsity models.
%as prior to the image reconstruction problem. 
Section IV provides the details of the developed image reconstruction algorithms
%ADMM algorithms that are used in inversion
%that solve the inverse problems 
and explains their efficient implementation
for cases with correlations all through the image data %Section V briefly describes how to adapt these algorithms to the case with correlations only in certain dimensions. 
or only in certain dimensions.
Numerical simulation results for three-dimensional reconstruction problems in computational spectral imaging are presented in Section V. Section VI concludes the paper and discusses the future directions.

%The performance of the developed algorithms is investigated for an application for a three-dimensional reconstruction problem in computational spectral imaging. Numerical results illustrate successful reconstructions.

%Developed methods are presented for convolutional image reconstruction problems with three-dimensional image of interest. Generally, a 3D signal of interest is correlated in all directions, which requires the use of 3D transforms or dictionaries in these models. However, in some applications, there may be no correlation in the third dimension. Therefore, 2D transforms or dictionaries are sufficient. In this paper, we develop algorithms for both cases.

%The focus is on three-dimensional problems, but developed algorithms can be easily extended to higher-dimensional cases. Moreover, convolutional inverse problems with three-dimensional image of interest are classified into two categories with three and two-dimensional correlations. Different analysis and synthesis priors are exploited for regularization, which are sparsifying transform, patch-based dictionary, and convolutional dictionary. The regularized problems are solved via the alternating direction method of multipliers (ADMM) exploiting efficient computation techniques~\cite{boyd2011}. Through numerical investigations, it has been shown that each prior has different merits or drawbacks.

\section{Forward Problem} %Forward Model
%linear system operator approximated by a linear combination of regular convolution operators.
In many image reconstruction problems in multidimensional imaging, the measurements can be modeled in the following form of superimposed convolutions:  
%in the following general convolutional form: 
\be \label{forward model}
%\begin{aligned}
    {y_{k}[n_1,n_2]}= \sum_{s=1}^{S}{x_{s}[n_1,n_2] * h_{k,s}[n_1,n_2]} + {w_{k}[n_1,n_2]},
%\end{aligned} 
\ee
where the measurement index $k$ = $1,\hdots,K$. 
This is a general multiple-input multiple-output model, where
${y_k}$'s denote the different 2D measurements and ${x_s}$'s represent the 2D slices of the unknown image %data
to be reconstructed. Hence, each measurement ${y_k}$ consists of blurred and superimposed images of ${x_s}$'s. We assume that the size of the measurements and the slices of the unknown image are both limited to $N \times N$.
Here, ${h_{k,s}}$  denotes the blur function (point-spread function) acting on the $s$th image slice, ${x_s}$, in the $k$th measurement, ${y_k}$. These blur functions generally model the slowly varying response of a shift-variant imaging system across the dimensions of time, space, spectral, depth, and such.
%Here, for $S > 1$,  generally $x_s$ represents the $sth$ slice of the datacube, and the measurements, $y_k$'s, are in the form of superimposed convolutions of the blurs, ${h_{k,s}}$, with these 2D slices. For example, in spectral imaging, $x_s$ corresponds to $sth$ spectral slice extracted from the spectral datacube, and hence the measurements, $y_k$, are in the form of superimposed convolutions of blurs, ${h_{k,s}}$, with spectral slices. %In depth imaging, each $x_s$ represents $sth$ depth slice, and the measurements,$y_k$, are in the form of superimposed convolutions of blurs with depth slices.  

This general forward model involving sum of convolutions is %a general form 
encountered in many imaging problems such as different three-dimensional image reconstruction problems ($K \geq 1$, $S \geq 1$) in computational imaging~\cite{ sroubek2016, denis2015, camera2015, zhang2010, badali2017, rau2011, stewart2011, kim2012,gol2016, florea2018, besson2019}, %youcef2018, mboula2016,
as well as classical and multiframe image deconvolution ($K \geq 1$, $S=1$). %~\cite{camera2015}. %\cite{  zhang2011} 
% different three-dimensional image reconstruction 
Note that our model allows each blurring operator to have a different weight as commonly used in the literature; for simplicity, these weights are simply embedded into the terms ${h_{k,s}}$'s in our model.

%photography\cite{denis2011,denis2015}, wave-field astronomical imaging\cite{denis2011,denis2015,camera2015}, three-dimensional microscopy\cite{preza2004,zhang2010,antipa2018}, spectral imaging~\cite{oktem2014,kamaci2017,kar2019, youcef2018}, ultrafast imaging\cite{badali2017}, radio interferometric imaging~\cite{stewart2011}, magnetic resonance imaging\cite{kim2012,gol2015}, and ultrasound imaging~\cite{florea2018,besson2019} ($K \geq 1$, $S>1$).

%ICIP2014 ve KAR2019'un union'u

Using lexicographic ordering and linearity of the convolution operator, the model in Eq.~\eqref{forward model} can be cast in the following matrix-vector form:
\be \label{forward model2}
\begin{aligned}
    &\hb{y}_{k} = \sum_{s=1}^{S}\hb{H}_{k,s}\hb{x}_{s} + \hb{w}_{k}.
\end{aligned} 
\ee
%where $\hb{H}_{k,s}$ is an $N^2 \times N^2$ Toeplitz matrix corresponding to the convolution operator with $\hb{h}_{k,s}$. 
Here, $\hb{y}_k \in \mathbb{R}^{N^2}$ represents %is lexicographically ordered
the noisy ${k}$th measurement vector, $\hb{x}_s \in \mathbb{R}^{N^2}$ denotes the vector for the $s$th image slice, and $\hb{H}_{k,s}\in \mathbb{R}^{N^2 \times N^2}$ is the convolution matrix representing %the convolution operation
the convolution with the blur function ${h_{k,s}}$. %is a convolution matrix which corresponds to the convolution operation with ${h_{k,s}}[n_1,n_2]$.
Lastly, $\hb{w}_k$ denotes the %additive 
white Gaussian noise vector 
%where $\hb{w}_k  \sim N(0,\sigma^2_k)$.  
%with mean zero and variance
whose each entry has mean zero and variance $\sigma^2_k$. 
By concatenating the measurement vectors $\hb{y}_k$'s and the image slice vectors $\hb{x}_s$'s vertically, the model in Eq.~\eqref{forward model2} can be expressed in the following final form:
%Let the blur function $h_{k,s}[n_1,n_2]$ has $L \times L$ support, i.e. $h_{k,s}[n_1,n_2] = 0$ for $n_1,n_2 \notin [-L/2, L/2-1]$. Assuming that the size of objects being imaged is limited to a smaller region than the input image size, i.e. ${x_s} = 0$ for $n_1,n_2 \notin [-(N-L + 1)/2, (N-L + 1)/2-1]$. Together, the linear convolution operator included in Eq.~\eqref{forward model} is replaced by a circular convolution of $N$ points, and hence the two-dimensional convolution operator can be represented as multiplication of a circular convolution matrix and a vectorized image. With this, the following matrix-vector form is obtained from the above image-formation model as
\be \label{forwardModel}
\begin{aligned}
    \hb{y}=\hb{H}\hb{{x}}+\hb{w},
\end{aligned} 
\ee
\be 
\hb{H}=\left( \begin{array}{ccc}
\hb{H}_{1,1} & \cdots & \hb{H}_{1, S} \\ 
\vdots & \ddots & \vdots \\ 
\hb{H}_{K,1} & \cdots  & \hb{H}_{K,S} \end{array}
\right),  
\hb{y}=\left[ \begin{array}{c}
\hb{y}_1 \\ 
{:} \\ 
\hb{y}_K\end{array}
\right], 
\hb{x}=\left[ \begin{array}{c}
\hb{x}_1 \\ 
{:} \\ 
\hb{x}_S\end{array}
\right] \nonumber
\ee
Here  % $\hb{y}_k \in \mathbb{R}^{N^2}$ is lexicographically ordered noisy ${k}$th measurement vector, 
$\hb{y} \in \mathbb{R}^{KN^2}$ is the vertically concatenated measurement vector. Similarly, % $\hb{x}_s \in \mathbb{R}^{N^2}$ denotes $s$th unknown image and 
the vector $\hb{x} \in \mathbb{R}^{SN^2}$ is the concatenated image vector, which consists of the image slices. The measurement matrix $\hb{H} \in \mathbb{R}^{KN^2 \times SN^2}$ contains all the convolution matrices involved. Lastly, the vector $\hb{w}$ = $[ \hb{w}_1^T \hdots \hb{w}_K^T ]^T$ denotes the overall noise vector. %Note that in implementation stage, these matrices and vectors are not formed; but, vertically concatenated 2D spatial slices and blur filters are utilized.

\section{Convolutional Inverse Problem}

In the inverse problem, the goal is to recover the unknown image vector, $\hb{x}$, 
%%%FIGEN: butun x ve y leri gozden gecirmelisin
from the noisy measurement vector, $\hb{y}$. Because each measurement is composed of blurred images of different slices, the overall task also involves the deconvolution of multiple objects.  
%related to the slices  which is related to the unknown $x$ through superimposed convolutions. 
This convolutional inverse problem is inherently ill-posed,
and as the blur functions, $h_{k,s}$, for different slices, $s$, or different measurements, $k$, become similar, %the problem becomes more ill-conditioned 
the conditioning of the problem gets worse due to the increase in the linear dependency of the columns or rows of $\hb{H}$, respectively.

%Here, the image data is reconstructed by combining information from multiplexed measurements with the additional prior (statistical or structural) knowledge about the unknown image.  %This ill-posed problem can be viewed as a type of multi-frame deconvolution problem involving multiple images. 
To incorporate the additional prior knowledge about the unknown image vector, this ill-posed inverse problem can be formulated as the following optimization problem:
\be
\label{rls}
\min_{{\hb{x}}} \; \frac{\beta}{2}|| \hb{y} - \hb{H} \hb{{x}} ||_2^2 + {\cal{R}}(\hb{{x}}),
\ee
where ${\cal{R}}(\hb{{x}})$ is the regularization functional.
This regularized least squares problem can also
be viewed as a maximum posterior estimation (MAP) problem that exploits statistical priors. Here, the first term controls data fidelity, whereas the second term 
controls how well the reconstruction matches our prior knowledge
of the solution, with the scalar parameter $\beta$ trading off
between these two terms.

For regularization, here sparse models in analysis or synthesis forms~\cite{elad2007} are exploited. 
In the analysis case, multidimensional discrete derivative operators or sparsifying transforms such as discrete cosine transform (DCT), wavelets, or their %high-dimensional 
Kronecker-product forms can be utilized~\cite{antipa2018, kar2019}. These generally yield fast and efficient computations. %implementations.
% duarte2012: Kronecker Compressive Sensing
In the synthesis case, multidimensional dictionaries can be utilized with local or global %dictionary 
sparsity models. One common approach is to partition the image data into patches and represent each local patch in terms of the elements of a small-size dictionary~\cite{aharon2006,ravishankar2010, caballero2014}. 
An alternative approach 
% To mitigate the partitioning problem, 
is to represent the entire image data with a global convolutional dictionary~\cite{wohlberg2016}. But there has been limited work for the three or higher dimensional convolutional dictionaries and their performance~\cite{nguyen2019, barajas2020}. %solano2019. 

Moreover, a multidimensional signal of interest may or may not be correlated in all directions, which %accordingly
consequently determines the dimension %size
of the transform or dictionary used in these sparse models (for example, 3D or 2D). Hence we can also adapt these models to correlations in different dimensions. Here, we formulate the inverse problem for all of these cases.

\subsection{Analysis Prior}

%\subsubsection{Analysis Model OR CASE ??} 
%In this case, %discrete derivative operators or sparsifying transforms can be utilized. Here, 
The analysis prior formulation is given by
\be
\label{analysis}
 {\cal{R}}(\hb{{x}}) = \hb{\Phi}(\bf{Tx}),
\ee
where $\bf{T}$ is a matrix representing an analysis operator such as a multidimensional transform or discrete derivative. %represents a fixed sparsifying transform matrix.
For example, for a 3D reconstruction problem, this analysis operator can be a 3D transform if there is correlation all through the image data, or if there is no correlation along the third dimension, the operator can perform a 2D transform along each slice.
%For various cases with or without correlations along the third dimension of the image data, analysis model in Eq.~\eqref{analysis} may include a single 3D or $S$ 2D discrete derivative operators or transforms. %with respect to the image data with or without correlations along the third dimension.
For the choice of regularization functional $\hb{\Phi}(.)$, there are different choices as well. 
%There are popular and powerful choices of the regularizer $\hb{\Phi}(.)$. 
One popular choice is $l_1$-norm, i.e. $\hb{\Phi}(\hb{Tx}) = ||\hb{Tx}||_1$.
%whose solution can be obtained through convex optimization techniques. 
Another %powerful
common choice is isotropic TV~\cite{lou2015}, i.e. $\hb{\Phi}(\hb{Tx})= \rm{TV}(\hb{x})$ with $\hb{T}$ = $\hb{I}$.
%%%TODO: Burada isotropic TV tanimi verilmeli diye dusunuyorum, ama artik buna takilmiyorum. Ama en azindan bu tanimin yer aldigi bir makaleye referans ekleyelim.
%which requires iterative method for the solution~\cite{chambolle2004}. 

%\subsubsection{Synthesis Prior}
\subsection{Synthesis Prior}

% and $\bf{P} = \bf{I}$ for convolutional dictionary.
In the synthesis case, the signal $\bf{x}$ is sparsely represented as a linear combination of %small number of 
columns (i.e. atoms) from a %synthesis 
dictionary $\bf{\tilde{D}}$. In general, the synthesis prior can be formulated as
%as $\bf{Dz} = \bf{x}$, where $\bf{D}$ denotes the dictionary matrix and $\bf{z}$ represents the sparse code. %\cite{elad2007}.
%weighted sum of atoms
%Hence the synthesis prior formulation can be given as %has the following generalized expression as
%\be
%\label{synthesis1}
%{\cal{R}}(\hb{{x}}) = \lambda||\bf{z}||_1 \; \text{s.t.} \quad \bf{Dz} = \bf{x}.
%\ee
%This representation requires inserting $\bf{Dz}$ into the data-fidelity term in Eq.~\eqref{rls} as $||\hb{y} - \hb{H} \hb{Dz} ||_2^2$, and causes a huge computational cost. In this work, to eliminate the computation of $\hb{H} \hb{D}$, we allow sparse approximation as $\bf{Dz} \approx \bf{Px}$ and represent the synthesis model as
\be \label{synthesis2}
\begin{aligned}
    &{\cal{R}}(\hb{{x}}) =  \underset{\hb{z}}{\min} \;
   \frac{1}{2} ||\bf{\tilde{D}} \hb{z}-\hb{Px}||^2_2+\lambda||\hb{z}||_1,
\end{aligned} 
\ee
where $\bf{z}$ represents the sparse code. The matrix $\hb{P}$ can be either identity or a patch operator depending whether the entire image or its patches are %sparsely 
represented.
%, or a patch operator when the sparse representation is for patches. 
%.extracts patches from the image data or is equal to $\hb{I}$. 
In this work, we consider both patch-based and convolutional dictionaries, %for the synthesis dictionary $\hb{D}$, 
which can be learned offline from a training data or online from the measurements through a dictionary update step. 

%Note that the Eq.~\eqref{synthesis2} may also include minimization with respect to the $\hb{D}$. This case is investigated in \ref{sec4}.
\subsubsection{Patch-based Dictionary}
Commonly, for efficient computation, a small-size dictionary is used for sparse representation of overlapping image patches. %of the image data 
 %in terms of the elements of 
This results in the following prior:
%For the reconstructed image $\hb{x}$ of large size, %the size of the reconstructed $\hb{x}$ is large, 
%the image data is divided into overlapping patches to reduce huge matrix-vector multiplications. Here, the synthesis model in Eq.~\eqref{synthesis2} is decomposed as % generalized and simplified expression of 
%the prior with the patch-based dictionary. For the image data with three-dimensional correlations, the synthesis model %with patch-based dictionary
%is expressed as % decomposed as
%% Hocam ben normalde buradaki patchleri tüm datada hem spatial hem de spectral domainde kayan patchler alacak şekilde extract ediyordum. Formulasyonları da ona göre yapmıştım. Örneğin patch boyutlatım n^2p ve image data boyutum N^2S idi. Ve toplamda N^2S adet patch extract ediyordum.
%%Fakat en son sadece spatial olarak kaydırarak alıp spectral olaraksa kaydırmadan patchleri extract etmeye ve tüm bantları almaya karar verdik. Şu makaleden yararlanmıştık.
%https://www.dropbox.com/s/xvvs0paxl7cgplu/brady2016_computationalSnapshotMultispeectral.pdf?dl=0
%Bu durumda ise image data boyutu N^2S olan bir datadan n^2S boyutunda toplam N^2 tane patch extract ediyorum. 
%Bu nedenle burdaki modelde küçük değişiklikler yapıyorum. Aynı değişikleri çözüm yöntemine de yansıtıyorum. Sadece birkaç index ve size için değişim yapıyorum ve computational cost'a da yansıtıyorum.

\be
\label{patch_based}
\begin{aligned}
    \underset{\hb{z}_{j}}{\min} \;  %\sum_{j=1}^{\bar{N}^2\bar{S}}
    \sum_{j} \frac{1}{2}||\hb{D}\hb{z}_{j}-\hb{P}_j\hb{x}||^2_2+\lambda||\hb{z}_{j}||_1, \\ 
\end{aligned} 
\ee
where $Q \times Q \times T$ is the patch size and $\hb{P}_j\in\mathbb{R}^{Q^2 T \times N^2S}$ is the matrix to extract the $j$th patch with $j=1,\hdots,J$. %are extracted from the image data of size $N \times N \times S$ by using the patch-extractor matrix . 
%Note that, we obtain $\bar{N}^2\bar{S}$ patches in total. For a special case with $s$ = $S$ and patches %extracted 
%with unity stride only in spatial dimensions, $\bar{N} = {N}$ and $\bar{S} = 1$ hold. % $N^2S$ patches are extracted with unity stride in all dimensions.
Moreover, $\hb{z}_j \in \mathbb{R}^{Q^2 T}$ denotes the sparse code %representation 
of the $j$th patch and $\hb{D} \in \mathbb{R}^{Q^2 T \times Q^2 T}$ represents the common %local 
dictionary used for all patches. 
%%%REVISION: Asagidaki aciklamanin derivasyonlarda herhangibir yerde kullanildigini gormedim, cikabilir.
Note that this prior is a special case of the general form given in Eq. \eqref{synthesis2} where now $\hb{P} = [\hb{P}_1^T|\hdots|\hb{P}_J^T ]^T$ and $\hb{\tilde{D}} = \hb{I}_J \Motimes \hb{D}$ with $\Motimes$ denoting the Kronecker product and $\hb{I}_n$ denoting an identity matrix of size $n\times n$.
%Here, the synthesis model in Eq.~\eqref{synthesis2} can be derived from Eq.~\eqref{patch_based} using the following relations as $\hb{z}$ = $[\hb{z}_1^T  \hdots \hb{z}_{N^2}^T ]^T$, $\hb{P}$ = $[\hb{P}_1^T  \hdots \hb{P}_{N^2}^T ]^T$. 
%$\hb{z}=\left[\begin{array}{c}
%\hb{z}_1 \\ 
%{:} \\ 
%\hb{z}_{N^2}\end{array}
%\right]$, $\hb{P} = \left[\begin{array}{c}
%\hb{P}_{1} \\ 
%{:} \\ 
%\hb{P}_{N^2}\end{array} 
%\right]$ 
%and $\hb{D}$ = $\hb{I}_{N^2}\Motimes\hb{D}$.
To adaptively learn the dictionary from the measurements, the prior in Eq.~\eqref{patch_based} can also be modified as follows: % minimized with respect to the dictionary as
\be
\label{patch_based_dic}
    \underset{\hb{z}_j,\hb{D}}{\min} \; \sum_{j}\frac{1}{2}||\hb{D}\hb{z}_{j}-\hb{P}_j\hb{x}||^2_2+\lambda||\hb{z}_{j}||_1 \; {\text{s.t.}} \; ||\hb{D}||_F = 1.
\ee
%\be
%\label{patch_based_dic}
%    \underset{\hb{z}_j,\hb{D}}{\min} \; \sum_{j=1}^{\bar{N}^2\bar{S}}\frac{1}{2}||\hb{D}\hb{z}_{j}-\hb{P}_j\hb{x}||^2_2+\lambda||\hb{z}_{j}||_1 \; {\text{s.t.}} \; ||\hb{D}||_2 = 1.
%\ee
%which results in additional dictionary updates. 
Here, the constraint %on the norm of local dictionary $\hb{D}$ 
is %required
needed to avoid the scaling ambiguity of %between 
the dictionary. % and sparse codes.

For an image data with two-dimensional correlations only, the patches can be extracted from each %$s$th 
image slice, $\hb{x}_s$. In this case,  there will be an additional summation over image slices %summed for $S$ image slices.
in Eq.~\eqref{patch_based} and~\eqref{patch_based_dic} with sparse code $\hb{z}_{j,s} \in \mathbb{R}^{Q^2}$ 
for the $j$th patch of the $s$th image slice, patch extraction matrix $\hb{P}_j\in\mathbb{R}^{Q^2\times N^2}$,  dictionary $\hb{D} \in \mathbb{R}^{Q^2 \times Q^2}$, and patch size $Q \times Q$. 
%Note that $\hb{P}$ and $\hb{\tilde{D}}$ in Eq. \eqref{synthesis2} are given by $\hb{P} = \hb{I}_S \Motimes \; [\hb{P}_1^T|\hdots|\hb{P}_J^T ]^T$ and $\hb{\tilde{D}} = \hb{I}_{SJ} \Motimes \hb{D}$.

%of size $n \times n$ are extracted from each $s$th image slice $\hb{x}_s$, by using the patch-extractor matrix $\hb{P}_j\in\mathbb{R}^{n^2\times N^2}$.  
%Note that we obtain $\bar{N}^2 = N^2$ patches with unity stride. 
%Here, $\hb{z}_j$ in Eq.~\eqref{patch_based} is replaced by $\hb{z}_{j,s} \in \mathbb{R}^{n^2}$ which denotes the sparse code %representation of $j$th patch for $s$th image slice and $\hb{D} \in \mathbb{R}^{n^2 \times n^2}$represents local dictionary which is commonly used by all patches. Finally,

%\be
%\label{patch_based}
%\begin{aligned}
%    \underset{\hb{z}_{j}}{\min} \; \frac{1}{2} \sum_{j=1}^{N^2S}(||\hb{D}\hb{z}_{j}-\hb{P}_j\hb{x}||^2_2+\lambda||\hb{z}_{j}||_1). \\ 
%\end{aligned} 
%\ee
%The synthesis model in Eq.~\eqref{synthesis2} can be derived from Eq.~\eqref{patch_based} using the following equations as $\hb{D}$ = $\hb{I}_{N^2S}\Motimes\hb{D}$, $\hb{z}=\left[\begin{array}{c}
%\hb{z}_1 \\ 
%{:} \\ 
%\hb{z}_{N^2S}\end{array}
%\right]$, and $\hb{P} = \left[\begin{array}{c}
%\hb{P}_{1} \\ 
%{:} \\ 
%\hb{P}_{N^2S}\end{array} 
%\right]$.
%Here, $\hb{P}_j\in\mathbb{R}^{n^2p\times N^2S}$ extracts the vectorized patch of size $n^2p$ from the vectorized image $\hb{x}$ of size $N^2S$. Also, $\hb{z}_j$ denotes sparse code %representation 
%of $j$th patch and $\hb{D}$ represents local dictionary which is commonly used by all $N^2S$ patches.
\subsubsection{Convolutional Dictionary} 
Alternatively, the entire image data can be represented using a global convolutional dictionary~\cite{wohlberg2016}, hence as a sum of three-dimensional convolutions with a few dictionary filters. This results in the following formulation:
%"which models an entire signal or image as a sum over a set of convolutions of coefficient maps, of the same size as the signal or image, with their corresponding dictionary filters."
\be \label{conv_dic}
\begin{aligned}
    & \underset{\hb{z}_{m}}{\min} \;
    \frac{1}{2} ||\sum_{m}\hb{d}_m * \hb{z}_m-\hb{x}||^2_2+\lambda \sum_{m}||\hb{z}_{m}||_1.\\
\end{aligned} 
\ee
where $\hb{d}_m \in \mathbb{R}^{L^2R}$ is the %vectorized 
$m$th dictionary filter of size $L \times L \times R$ and $\hb{z}_m \in \mathbb{R}^{N^2S}$ is the corresponding %vectorized 
sparse code of size $N \times N \times S$ with $m=1,\hdots,M$. In this representation, the dictionary size is generally much smaller than the image size (i.e. $L \ll N$ and $R \ll S$), while the sparse code is of the same size as the image.
%to provide the circular boundary conditions~\cite{wohlberg2016}. 
Also note that the dictionary $\hb{\tilde{D}}$ in Eq.~\eqref{synthesis2} has a specific form in this case as given by $\hb{\tilde{D}} = [\hb{D}_1  \hdots \hb{D}_M ]$ with $\hb{D}_m $ representing the %circular 
convolution matrix for the $m$th dictionary filter and %$\hb{P} = \hb{I}_{N^2S}$.
$\hb{P}$ is identity as the entire image is represented.
%%%REVISION: Yukarida "circular" vurgusu olabilir, wohlberg kendi makalesinde vurguluyor.
%By inserting, $\hb{P}$ = $\hb{I}$, $\hb{z}$ = $[\hb{z}_1^T  \hdots \hb{z}_M^T ]^T$ and $\hb{D}$ = [$\hb{D}_1 $ \hdots $\hb{D}_M $] where $\hb{D}_m $ is circular convolution matrix of dictionary filter $\hb{d}_m$, generalized synthesis model in Eq.~\eqref{synthesis2} is obtained. 
As before, %the prior with synthesis model in Eq.~\eqref{conv_dic} is minimized with respect to the convolutional dictionary as
to adaptively learn the dictionary from the measurements, the prior in Eq.~\eqref{conv_dic} can also be modified as % minimized with respect to convolutional dictionary as
\begin{eqnarray}
  \label{DL_Prior1}
    & \underset{\hb{z}_m,\hb{d}_m}{\min} & \;
   \frac{1}{2}||\sum_{m}\hb{d}_m * \hb{z}_{m}-\hb{x}||^2_2+\lambda \sum_{m}||\hb{z}_{m}||_1 \\ \nonumber
    & {\text{s.t.}} & \quad||\hb{d}_m||_2 = 1 \quad m = 1,2,\hdots, M.
\end{eqnarray}
%which results in additional dictionary updates. 

For an image data with two-dimensional correlations only, a common convolutional dictionary can be used to represent 
each image slice, $\hb{x}_s$, separately. In this case,  there will be an additional summation over image slices %summed for $S$ image slices.
in Eq.~\eqref{conv_dic} and \eqref{DL_Prior1} with the $m$th dictionary filter $\hb{d}_{m} \in \mathbb{R}^{L^2}$ of size $L \times L$ and the corresponding sparse code $\hb{z}_{m,s} \in \mathbb{R}^{N^2}$ 
for the $s$th image slice. 
%Here the dictionary $\hb{\tilde{D}}$ in Eq.~\eqref{synthesis2} has a specific form in this case as given by $\hb{\tilde{D}} = \hb{I}_S \Motimes \; [\hb{D}_1| \hdots|\hb{D}_M ]$ and $\hb{P} = \hb{I}_{N^2S}$.

%each image slice $\hb{x}_s$ expressed as sum of convolutions, which adds an index $s$ to the variables. Therefore, $\hb{d}_{m,s} \in \mathbb{R}^{L^2}$ is the vectorized $m$th dictionary filter of size $L \times L$ and $\hb{z}_{m,s} \in \mathbb{R}^{N^2}$ is the corresponding vectorized sparse code of size $N \times N$ for the image slice $\hb{x}_s$. Finally, the expression in Eq.~\eqref{conv_dic} or~\eqref{DL_Prior1} is summed for $S$ image slices.

\subsubsection{Convolutional Dictionary with Tikhonov Regularization}

Because convolutional dictionaries are known to work well for the representation of high-frequency components of a signal, they are commonly used after highpass filtering~\cite{wohlberg2016, venkatakrishnan2019}. An alternative to this preprocessing is to introduce gradient (Tikhonov) regularization to the sparse code $\hb{z}_m$'s as follows~\cite{wohlberg20162}:
\begin{eqnarray}
 \label{DL_PriorHP_conv}
%\begin{aligned}
 \underset{\hb{z}_{m}}{\min} \; \frac{1}{2}
    ||\sum_{m}\hb{d}_m * \hb{z}_{m}-\hb{x}||^2_2+\lambda \sum_{m}||\hb{z}_{m}||_1 + \\ \nonumber  \frac{\mu}{2} \sum_{m}||\hb{r}_1*\hb{z}_{m}||^2_2 + ||\hb{r}_2*\hb{z}_{m}||^2_2 + ||\hb{r}_3*\hb{z}_{m}||^2_2,
%\end{aligned} 
\end{eqnarray}
where $\hb{r}_1$, $\hb{r}_2$ and $\hb{r}_3$ are respectively the filters that compute the gradient along the first, second and third dimensions.
% " convolutional sparse coding/dictionary learning algorithms are applied to test/training images after a highpass filtering preprocessing step. representation of the highpass component, and a complete image representation includes both the lowpass component and the sparse representation"
% "Since convolutional sparse representations do not provide a good representation of the low-frequency components of an image, it is common practice to compute the representation from a contrastnormalized[8] or highpass filtered [3] version of the image to be decomposed.
The reasoning behind this regularization is based on the following observation. 
%that for the highpass component of a signal to have a good convolutional sparse representation is equivalent to the overall signal
Note that if the highpass component of a signal has a good convolutional representation, i. e. $\hb{x}_{h} = \sum_{m} \hb{d}_m * \hb{z}_m$, then the overall signal also has the representation $\hb{x} = %\hb{g}^{-1} *\hb{x}_{h} =%
\sum_{m} \hb{d}_m *(\hb{g}^{-1} * \hb{z}_m)$, where $\hb{g}^{-1}$ is the inverse of the filter that computes the highpass component. Hence, equivalently, low-pass filtered $\hb{z}_m$'s, or Tikhonov regularized $\hb{z}_m$'s, can provide a good convolutional representation for the overall signal.   

%Assume we represent the high-frequency component as a sum of convolutions  $\hb{x}_{h} = \sum_{m = 1}^{M} \hb{d}_m * \hb{z}_m = \hb{g} * \hb{x} $
%\be \label{intuition}
%\begin{aligned}
%   \hb{x}_{HPF} = \hb{g} * \hb{x}  = \sum_{m = 1}^{M} \hb{d}_m * \hb{z}_m 
%\end{aligned} 
%\ee
%where $\hb{g}$ represents an invertible FIR high-pass filter. By convolving right and left hand sides with $\hb{g}^{-1}$ we obtain the original image $\hb{x}$ as
%\be \label{intuition2}
%\begin{aligned}
%    \hb{x} = \hb{g}^{-1} *\hb{x}_{h} =  \sum_{m = 1}^{M} \hb{d}_m *(\hb{g}^{-1} * \hb{z}_m).
%\end{aligned} 
%\ee
%Here, this operation corresponds to convolving  $\hb{z}_m$'s with a low-pass filter. To acquire low-pass filtered $\hb{z}_m$'s, smoothing with Tikhonov  regularization has been considered as an effective approach~\cite{wohlberg20162}.  

 The above formulation can be similarly modified to learn the dictionary adaptively from the measurements. % similar to Eq.~\eqref{DL_Prior1}. Note that,
 Moreover, for an image data with two-dimensional correlations, only the gradients along the first and second directions are needed for the regularization. %, and sum the expression in Eq.~\eqref{DL_PriorHP_conv} for the $S$ image slices.

% === IV. Image Recontruction Methods ========================================
% =================================================================================
\section{Image Reconstruction Algorithms}

%In this section, developed methods are presented for convolutional image reconstruction problems with three-dimensional image of interest. Generally, a 3D signal of interest is correlated in all directions, which requires the use of 3D transforms or dictionaries in these models. However, in some applications, there may be no correlation in the third dimension~\cite{oktem2014, kamaci2017, dogan2019}. Therefore, 2D transforms or dictionaries are sufficient. In this paper, we develop algorithms for both cases. The focus is on three-dimensional problems, but developed algorithms can be easily extended to higher-dimensional cases. 

% Moreover, convolutional inverse problems with three-dimensional image of interest are classified into two categories with three and two-dimensional correlations. Different analysis and synthesis priors are exploited for regularization, which are sparsifying transform, patch-based dictionary, and convolutional dictionary. %The regularized problems are solved via the alternating direction method of multipliers (ADMM) exploiting efficient computation techniques~\cite{Boyd2011}. %Through numerical investigations, it has been shown that each prior has different merits or drawbacks.

%The resulting optimization problem is solved for the unknown image $\hb{x}$ for analysis prior. It is solved for the unknown image $\hb{x}$, sparse code $\hb{z}$ (and dictionary $\hb{D}$ with online dictionary learning) for synthesis prior by using Alternating Direction Method of multipliers (ADMM)~\cite{boyd2011}. 

We now focus on developing efficient algorithms for solving the resulting optimization problems. 
Based on alternating direction method of multipliers (ADMM)~\cite{boyd2011,afonso2010}, we present reconstruction algorithms with closed-form and efficient update steps for both analysis and synthesis cases. %ADMM is an optimization algorithm that is used in many signal and image reconstruction problems~\cite{solano2019, nguyen2018, wetzstein2018, cardona2018}.
%ADMM solves distributed unconstrained optimization problems by splitting them into sub-problems, which are solved alternatingly\cite{solano2019, nguyen2018, wetzstein2018, cardona2018}. In this section, the developed methods are presented for convolutional image reconstruction problems with three-dimensional image of interest. The solutions are presented for the case with correlation along the third dimension for both analysis and synthesis models.

\subsection{Analysis Case}
%\subsection{Analysis Model}
Using the ADMM framework, variable splitting is applied to the regularized least-squares problem in Eq.~\eqref{rls} with the analysis prior in Eq.~\eqref{analysis}. This results in the following problem:
%In this part, we solve the regularized least-squares problem in Eq.~\eqref{rls} by inserting the prior information with analysis model in Eq.~\eqref{analysis} for the regularization term ${\cal{R}}(\hb{{x}})$. We solve this problem by developing a fast reconstruction algorithm that is based on ADMM. After variable-splitting, we arrive at the following problem: 
\be \label{analysis_ADMM}
\begin{aligned}
    &\underset{\hb{x,\,t}}{\min}\;\frac{\beta}{2}|| \hb{y} - \hb{H} \hb{{x}} ||_2^2 + \lambda \Phi(\hb{t}) \quad \text{s.t.}\quad \hb{t}=\hb{Tx},
\end{aligned} 
\ee
where $\hb{t}$ is the auxiliary variable in the ADMM framework. Expressing the problem in Eq.~\eqref{analysis_ADMM} in an augmented Lagrangian form~\cite{afonso2010} yields to a minimization over $\hb{x}$ and $\hb{t}$. We minimize over each in an alternating fashion as follows:
%This problem is solved with respect to variables $\hb{x}$ and $\hb{t}$ using the alternating minimization approach. %This problem is solved using the alternating minimization approach with respect to the variables $\hb{x}$ and $\hb{t}$. 
%ADMM steps for this alternating minimization process are expressed as
\be \label{iter_ADMM1}
%\begin{aligned}
    \hb{x}^{l+1} = \underset{\hb{x}}{\arg \min}\; \frac{\beta}{2}|| \hb{y} - \hb{H} \hb{{x}} ||_2^2 +\frac{\rho}{2}||\hb{Tx} - \hb{t}^l + \hb{u}^l||_2^2,
%\end{aligned} 
\ee
\be \label{iter_ADMM2}
\begin{aligned}
&    \hb{t}^{l+1} = \underset{\hb{t}}{\arg \min}\; \lambda \Phi(\hb{t})+\frac{\rho}{2}||\hb{Tx}^{l+1} - \hb{t} + \hb{u}^l||_2^2,
\end{aligned} 
\ee
\be \label{subu_trans}
\begin{aligned}
    \hb{u}^{l+1} = \hb{u}^l + \hb{Tx}^{l+1} - \hb{t}^{l+1},
\end{aligned} 
\ee
where $\rho$ is a penalty parameter and the last equation is for the update of the dual variable $\hb{u}$ in the ADMM framework.
%with $\hb{u}$ denoting the dual variable in the ADMM framework. % $\hb{u}$ is dual variable of $\hb{Tx}$ which is updated as in Eq.~\eqref{subu_trans}. %is its update. %the update of the dual variable.
The efficient solutions of the first two problems are explained in the image %$\hb{x}$ 
update and auxiliary variable %$\hb{t}$ 
update steps, respectively. 

%We explain how to efficiently solve each minimization problem above in the image %$\hb{x}$ 
%update and auxiliary variable %$\hb{t}$ 
%update steps.
\subsubsection{Image update}
The minimization in Eq.~\eqref{iter_ADMM1} over the image $\hb{x}$ 
%problem in Eq.~\eqref{iter_ADMM1} over the image $\hb{x}$ is 
corresponds to a least-squares problem with the following closed-form solution:
\be \label{solx_Trans}
\begin{aligned}
    \hb{x} = (  \rho \hb{T}^H\hb{T} + \beta \hb{H}^H\hb{H})^{-1}(\beta \hb{H}^H\hb{y} + \rho\hb{T}^H(\hb{t-u})).
\end{aligned} 
\ee
Here $\hb{T}$ is generally a unitary transform resulting in $\hb{T}^H\hb{T} = \hb{I}$. This solution can be efficiently obtained through computations in the frequency domain by exploiting the property of the 2D circular convolutions involved~\cite{kamaci2017}. % matrices. 
%utilizing the fact that circular convolution is diagonalizable by the DFT.

%%%REVISION: Asagidaki ÜÇ PARAGRAF bu makaledeki en yeni kisimlardan biri oldugu icin (literature gore), cok detayli aciklanmasi gerekiyordu...Su anki hali hala yeteri kadar anlasilir olmayabilir. 
Note that each block of $\hb{H}$ is diagonalized by the discrete Fourier transform (DFT) matrix since $\hb{H}_{k,s}$ is block circulant with circular blocks (BCCB). 
%"If matrix represents a 2-D cyclic convolution (periodic boundary conditions), it is a block circulant matrix with circulant blocks that can be factorized as, where is the unitary matrix representing the discrete Fourier transform (DFT) and is diagonal.
Hence $\hb{H}_{k,s} =  \hb{F}_{2D}^H \hb{\Lambda}_{k,s} \hb{F}_{2D}$ where $\hb{F}_{2D}$ %\in \mathbb{R}^{N^2 \times N^2}
is the unitary 2D DFT matrix and $\hb{\Lambda}_{k,s}$ is a diagonal matrix whose diagonal consists of the 2D DFT of the blur function $h_{k,s}$, with $k=1,\hdots,K$ and $s=1,\hdots, S$.
%the unitary matrix representing the 2D DFT matrix. 
As a result, the overall matrix $\hb{H}$ can be written as $\hb{H} =  \hb{\bar{F}}^H \hb{{\Lambda}} \hb{\tilde{F}}$ where $\hb{\bar{F}}  = \hb{I}_K \Motimes \hb{F}_{2D}$ and $\hb{\tilde{F}}  = \hb{I}_S \Motimes \hb{F}_{2D}$ with $\Motimes$ denoting the Kronecker product and $\hb{I}_n$ denoting an identity matrix of size $n\times n$. Here
$\hb{{\Lambda}}$ is a matrix of $K \times S$ blocks with each block given by $\hb{\Lambda}_{k,s}$. By inserting this expression of $\hb{H}$ in Eq.~\eqref{solx_Trans}, the following form can be obtained for the efficient computation of the image
update step:
\be
\label{solx_Trans_DFT}
    \hb{x} = \hb{\tilde{F}}^{H}(\rho \hb{I} + \beta \hb{{\Lambda}}^H\hb{{\Lambda}})^{-1}(\beta\hb{{\Lambda}}^H\hb{\bar{F}} \hb{y} +  \rho\hb{\tilde{F}}\hb{T}^H(\hb{t-u})).
\ee 

For the computation of Eq.~\eqref{solx_Trans_DFT}, forming any of the matrices is not required, which provides huge savings for the memory as well as the computation time. Here, multiplication by $\hb{\tilde{F}}$ or $\hb{\tilde{F}}^H$ corresponds to taking the Fourier or inverse Fourier transforms of all 2D slices for $s=1, \hdots, S$. %slices,
%respectively. 
Similarly, multiplication by $\hb{\bar{F}}$ corresponds to taking the Fourier transforms of all 2D measurements for $k=1, \hdots, K$.
For example, $\hb{\bar{F}}\hb{y}=[ (\hb{F}_{2D}\hb{y}_1)^T | \hdots | (\hb{F}_{2D}\hb{y}_K)^T]^T$, where each term can be computed via the 2D FFT. Moreover, because $\hb{{\Lambda}}$ is a block matrix consisting of diagonal matrices, multiplication by $\hb{{\Lambda}}^H$ corresponds to element-wise 2D multiplication with the conjugated DFTs of the underlying blur functions and summation. % For example, in $\hb{{\Lambda}}^H\hb{\bar{F}} \hb{y} =[ \hb{b}_1^T| \hdots | \hb{b}_S^T]^T$,  where  $\hb{b}_s =\sum_{k=1}^{K}\hb{\Lambda}_{k,s}^H\hb{F_{2D}}\hb{y}_k$, each $\hb{b}_s$ is computed as sum of element-wise multiplications. %( NOT: BIRAZ DAHA AÇIKLAYICI OLMASI AÇISINDAN BU CÜMLEYİ EKLEYEBİLİR MİYİM)
Furthermore, for a unitary $\hb{T}$, multiplication by $\hb{T}^H$ corresponds to taking the inverse transform.  %$t[n_1,n_2,s] - u[n_1,n_2,s]$ 
%of $\hb{t-u}$. 
Note that when the image data is correlated in all directions, this will be a 3D transform; otherwise, it will be a 2D transform applied on each slice separately.
%when the image data is correlated along the third dimension. For the image data with 2D correlations, $\hb{T}^H$ simply takes inverse 2D transform of each slices.

% \hb{I}_{N^2S}
Lastly, the inverse of $\hb{\Psi} = \rho \hb{I} + \beta \hb{{\Lambda}}^H\hb{{\Lambda}}$ needs to be computed only once, and hence does not affect the computational cost of the iterations. However, it is possible to reduce the required time and memory for this pre-computation through a recursive
block matrix inversion approach~\cite{noble1988}.
Note that $\hb{\Psi}$ is a block matrix of $S \times S$ blocks, where each block is a diagonal matrix  given by $\hb{\Psi}_{i,j} = {\delta}_{i,j}\hb{I} + \beta \sum_{k=1}^{K}\hb{{\Lambda}}_{i,k}^H\hb{{\Lambda}}_{k,j}$,
% \hb{I}_{N^2}
%\be \label{psi}
%\begin{aligned}
%    \hb{\Psi}_{i,j} = ({\delta}_{i,j}\hb{I} + \beta \sum_{k=1}^{K}\hb{{\Lambda}}_{i,k}^H\hb{{\Lambda}}_{k,j}),
%\end{aligned}
%\ee
with $\hb{\delta}_{i,j}$ denoting the Kronecker delta function and $i,j=1,\hdots,S$. Hence, for $S = 2$, the inverse can be computed as 
\be \label{recinv}
\begin{aligned}    
    \begin{bmatrix}
        \hb{\Psi}_{1,1} &  \hb{\Psi}_{1,2} \\  
        \hb{\Psi}_{2,1} &  \hb{\Psi}_{2,2} \end{bmatrix}^{-1} =  \begin{bmatrix} 
        \hb{A} &  \hb{\Psi^{-1}}_{1,1}\hb{\Psi}_{1,2}\hb{B} \\  
        \hb{B}\hb{\Psi}_{2,1}\hb{\Psi}_{1,1}^{-1} &  \hb{-B} \end{bmatrix},
\end{aligned} 
\ee
\noindent where $\hb{A}$ = $\hb{\Psi}_{1,1}^{-1} - \hb{\Psi}_{1,1}^{-1}\hb{\Psi}_{1,2}\hb{B}\hb{\Psi}_{2,1}\hb{\Psi}_{1,1}^{-1}$ and $\hb{B}$ = $- (\hb{\Psi}_{2,2} - \hb{\Psi}_{2,1}\hb{\Psi}_{1,1}^{-1}\hb{\Psi}_{1,2})^{-1}$. For $S > 2$, the matrix $\hb{\Psi}$ can be partitioned into $2 \times 2$ blocks and each block can be inverted recursively using Eq.~\eqref{recinv}.
Because all the matrices involved in these computations are diagonal, computing the inverse of $\hb{\Psi}$ requires simple element-wise 2D multiplication and division operations.

\subsubsection{Auxiliary variable update}
The minimization in Eq.~\eqref{iter_ADMM2} over the auxiliary variable $\hb{t}$ depends on the choice of the regularization functional $\hb{\Phi}(.)$. 
%problem in Eq.~\eqref{iter_ADMM1} over the image $\hb{x}$ is 
If $\hb{\Phi}(.)$ = $||.||_1$, the solution is given by a soft-thresholding operation:% Eq.~\eqref{iter_ADMM1} is solved via soft-thresholding as 
\be \label{solt_trans}
\begin{aligned}
    & \hb{t} = {\rm{soft}}(\hb{Tx} +\hb{u},\lambda/\rho). \\
\end{aligned} 
\ee
Here, multiplication by $\hb{T}$ corresponds to either a single 3D transform or multiple 2D transforms along each slice. Moreover,  the soft-thresholding operation ${\rm{soft}}(\hb{w}, \tau)$ is component-wise computed as $w_i \rightarrow {\rm{sign}}(w_i) \max(|w_i|- \tau, 0)$ for all $i$, where ${\rm{sign}}(w_i)$ takes value $1$ if $w_i > 0$ and $-1$ otherwise.
%component-wise soft-thresholding operation is performed as $\hb{t} = sign(\hb{Tx} +\hb{u}) \odot max(0, |\hb{Tx} +\hb{u}|- \lambda/\rho)$.  
%\be \label{solt_trans2}
%\begin{aligned}
%    & \hb{t} = sign(\hb{Tx} +\hb{u}) \odot max(0, |\hb{Tx} +\hb{u}|- \lambda/\rho), \\
%\end{aligned} 
%\ee
%which is an element-wise operation.
%This is an element-wise opeation and performed without forming the vectors.
If $\hb{\Phi}(.)$ is chosen as isotropic TV, the solution can be obtained in this case using Chambolle's algorithm~\cite{chambolle2004}.

\begin{algorithm}[t]
\caption{Image reconstruction algorithm: analysis case}
\label{alg:algorithm_sum1}
% \hspace*{\algorithmicindent} 
\textbf{Input:}           $\hb{y}$: measurement,
        $\hb{H}$: system matrix,
        $\hb{T}$: transform
% \hspace*{\algorithmicindent} 
\textbf{Output:}
$\hb{x}$: reconstructed image 
\begin{algorithmic}[1]
%        \Require
%        $\hb{y}$: measurement,
%        $\hb{H}$: system matrix,
%        $\hb{T}$: transform
%        \Ensure
%        $\hb{x}$: reconstructed image 
        \State Choose $\lambda > 0$, $\rho > 0$, $\beta > 0$  , $\hb{t}^0$, and $\hb{u}^0$.
\State  Compute $(\rho \hb{I} + \beta \hb{{\Lambda}}^H\hb{{\Lambda}})^{-1}$ and $\beta\hb{{\Lambda}}^H\hb{\bar{F}} \hb{y}$.
\Repeat
\State{update $\hb{x}^{l+1}$ using Eq.~\eqref{solx_Trans_DFT}}
\State{update $\hb{t}^{l+1}$ using Eq.~\eqref{iter_ADMM2}} or Eq.~\eqref{solt_trans} 
\State{update $\hb{u}^{l+1}$ as $\hb{u}^{l+1} = \hb{u}^l + \hb{Tx}^{l+1} - \hb{t}^{l+1}$}
\State{$l \leftarrow l+1$}
\Until{stopping criterion is satisfied.}
\end{algorithmic}
\end{algorithm}
%\begin{algorithm}[t]
%    \caption{Image Reconstruction Algorithm with Sparsifying Transform}
%    \label{alg:algorithm_sum1}
%    \begin{algorithmic}[1]
%        
%        \Require
%        $\hb{y}$: Measured image,
%        $\hb{H}$: Blur filter,
%        $\hb{T}$: Sparsifying Transform
%        \Ensure
%        $x$: Reconstructed Image 
%        \State Choose: $\lambda > 0$, $\rho > 0$, $\beta > 0$  , $\hb{t}^0$, $\hb{u}^0$
%        \State  $\hb{\Psi}^{-1}$ = $(\rho \hb{T}^H\hb{T} + \beta \hb{H}^H\hb{H})^{-1}$ is computed.  
%        \State $\hb{\Omega}$ = $\beta \hb{H}^H\hb{y}$ is computed
%        \Repeat
%        
%        \State{$\hb{x}^{l+1}$ = $\hb{\Psi}^{-1}(\hb{\Omega}+ \rho\hb{T}^H(\hb{t}^l-\hb{u}^l) )$}
%        \State{$\hb{t}^{l+1}$} = $\Psi_{\Phi}(\hb{Tx}^{l+1}-\hb{u}^l)$
%        \State{$\hb{u}^{l+1} = \hb{u}^l + \hb{Tx}^{l+1} - \hb{t}^{l+1}$}
%        \Until{Some convergence criterion is satisfied}
%    \end{algorithmic}
%\end{algorithm}

\subsection{Synthesis Case}
%\subsection{Synthesis Model}
\label{convalgorithm}

Using the ADMM framework, variable splitting is applied to the regularized least-squares problem in Eq.~\eqref{rls} with the synthesis prior in Eq.~\eqref{synthesis2}. This results in the following problem:
\be \label{synthesis_ADMM}
\begin{aligned}
    & \underset{\hb{x,\,z,\,t}}{\arg \min}  \; \frac{\beta}{2}|| \hb{y} - \hb{H} \hb{{x}} ||_2^2 +\frac{1}{2} ||\hb{\tilde{D}z} - \hb{Px}||_2 +\lambda ||\hb{t}||_1 & \\
    &\text{s.t.} \quad  \hb{t} = \hb{z}, &
\end{aligned} 
\ee
where $\hb{t}$ is the auxiliary variable in the ADMM framework. As discussed before, for patch-based and convolutional dictionary cases, the matrices $\hb{\tilde{D}}$ and $\hb{P}$ take particular forms. 
Expressing the problem in Eq.~\eqref{synthesis_ADMM} in an augmented Lagrangian form~\cite{afonso2010} yields to a minimization over $\hb{x}$, $\hb{z}$ and $\hb{t}$. We minimize over each in an alternating fashion as follows:
%. ADMM steps for this alternating minimization 
%process are expressed as
%This problem is alternatingly solved with respect to variables $\hb{x}$, $\hb{z}$ and $\hb{t}$ using the alternating minimization approach. ADMM steps for this %alternating minimization 
%process are expressed as%ADMM iterations for this problem are expressed as
\be \label{iterd_ADMM1}
\begin{aligned}
     \hb{x}^{l+1}= \underset{\hb{x}}{\arg \min}\;\frac{\beta}{2}|| \hb{y} - \hb{H} \hb{{x}} ||_2^2 +\frac{1}{2}||\hb{\tilde{D}} \hb{z}^{l} - \hb{Px}||_2,
\end{aligned} 
\ee
\be \label{iterd_ADMM2}
%\begin{aligned}
    \hb{z}^{l+1} = \underset{\hb{z}}{\arg \min}\; \frac{1}{2}||\hb{\tilde{D}z} - \hb{Px}^{l+1}||_2^2 + \frac{\rho}{2}||\hb{z} - \hb{t}^l + \hb{u}^l||_2^2,
%\end{aligned} 
\ee
\be \label{iterd_ADMM3}
\begin{aligned}
    &\hb{t}^{l+1} = \underset{\hb{t}}{\arg \min}\; \lambda||\hb{t}||_1 +||\hb{z}^{l+1} - \hb{t} + \hb{u}^l||_2^2,
\end{aligned} 
\ee
\be \label{subu_dic}
\begin{aligned}
    \hb{u}^{l+1} = \hb{u}^l + \hb{z}^{l+1} - \hb{t}^{l+1},
\end{aligned} 
\ee 
where the last equation is for the update of the dual variable $\hb{u}$ in the ADMM framework. The efficient solutions of the first three problems are explained in the image %$\hb{x}$ 
update, sparse code update and auxiliary variable %$\hb{t}$ 
update steps, respectively. 

%with $\hb{u}$ denoting the dual variable in the ADMM framework. %Here, $\hb{u}$ is dual variable of $\hb{t}$, which has the update in Eq.~\eqref{subu_dic}. %and Eq.~\eqref{subu_trans} is the update of the dual variable. 
%We now explain how to efficiently solve the problems in Eq.~\eqref{iterd_ADMM1},~\eqref{iterd_ADMM2} and~\eqref{iterd_ADMM3} which  we refer as image $\hb{x}$ update, sparse code $\hb{z}$ update and auxiliary variable $\hb{t}$ update.

%%%EGER KULLANILIYORLARSA, SECTION 3'TE TANIMLANMALI : When the image data is correlated along all directions, $\hb{P} = [\hb{P}_1^T|\hdots|\hb{P}_J^T ]^T$ and   $\hb{\tilde{D}} = \hb{I}_J \Motimes \hb{D}$ for patch-based dictionary where $J$ denotes the number of patches. Similarly, $\hb{P} = \hb{I}_S \Motimes \; [\hb{P}_1^T|\hdots|\hb{P}_J^T ]^T$ and $\hb{\tilde{D}} = \hb{I}_{SJ} \Motimes \hb{D}$ for the image data with 2D correlations only. Moreover, $\hb{P} = \hb{I}_{N^2S}$ for convolutional dictionaries. Here, when the there is correlations all through image data, $\hb{\tilde{D}} = [\hb{D}_1|\hdots|\hb{D}_M ]$ and otherwise $\hb{\tilde{D}}$ = $\hb{I}_S \Motimes \; [\hb{D}_1| \hdots|\hb{D}_M ]$. Note that, $\hb{z}$ denotes the vertically concatenated sparse codes for all cases. 

\subsubsection{Image update}
%For image update, we need to solve the problem in Eq.~\eqref{iterd_ADMM1}. Because image update is a least-squares problem, it has a closed-form solution as
%For  minimization  over $\hb{x}$, we face a least-squares problem which has the following closed-form solution:
The minimization in Eq.~\eqref{iterd_ADMM1} over the image $\hb{x}$ 
%problem in Eq.~\eqref{iter_ADMM1} over the image $\hb{x}$ is 
corresponds to a least-squares problem with the following closed-form solution:
\be \label{solx_synt}
\begin{aligned}
    \hb{x} = ( \hb{P}^H\hb{P}+ \beta \hb{H}^H\hb{H})^{-1}(\beta \hb{H}^H\hb{y} + \hb{P}^H\hb{\tilde{D}z}).
\end{aligned} 
\ee
%Considering the synthesis models in Eq.~\eqref{patch_based} and~\eqref{conv_dic},
Here $\hb{P^HP}$ %t\hb{I}_{N^2S}$
is a scaled identity matrix, i.e. $\hb{P^HP} = t\hb{I}$, where for the convolutional dictionary case $t=1$ and for the patch-based dictionary case the constant $t$ depends on the patch parameters~\cite{ravishankar2010}. 
%size and the number of the patches~\cite{addReference} 
%burda tyi etkileyen çok şey var. patch sayısı, boyutu, stride sayısı, zeropad edip etmediğimiz vs. daha önceden bilinen şeyler oldukları için detaylandırmadım.
As in the analysis case, this solution can be efficiently obtained through computations in the frequency domain.
%$Q^2T \hb{I}$ or % $\hb{P}$ = $\hb{I}$ and 
%$\hb{P}^H\hb{P}$ =
%$\hb{I}$ of size $N^2S \times N^2S$ for patch-based and convolutional dictionaries, respectively. %of size $N^2S$.
After similar steps, the
following form can be obtained for the efficient computation of the image update step:
\be
\label{solx_synt_DFT}
%\begin{aligned}
    \hb{x} = \hb{\tilde{F}}^H(t\hb{I} + \beta \hb{{\Lambda}}^H\hb{{\Lambda}})^{-1}(\beta\hb{{\Lambda}}^H\hb{\bar{F}}\hb{y} + \hb{\tilde{F}}\hb{P}^H\hb{\tilde{D}z}).
%\end{aligned} 
\ee 
This can be computed in a similar way as Eq.~\eqref{solx_Trans_DFT}, except the last term.

For the patch-based dictionary case, the last term $\hb{\tilde{F}}\hb{P}^H\hb{\tilde{D}z}$,
= $\sum_{j}\hb{\tilde{F}}\hb{P}_j^H\hb{D}\hb{z}_j$, %is computed in the following order: First, 
with $\hb{D}$ and $\hb{z}_j$ representing the common dictionary used for all patches and the resulting sparse codes, respectively.
%local dictionary matrix $\hb{D}$ and each sparse vector $\hb{z}_j$ is multiplied and results in a vector of size $Q^2T$ or $Q^2$ for the image data with 3D or 2D correlations, respectively. 
Here $\hb{P}_j^H$ is the adjoint of the patch-extraction operation with patch size $Q \times Q \times T$, and hence converts a vector of length $Q^2T$ to a 3D signal of size $N \times N \times S$. Lastly, multiplication by $\hb{\tilde{F}}$ corresponds to taking 2D FFTs along all $N\times N$ slices. % for $s=1,\hdots,S$. 
For an image data with 2D correlations only, the patches are extracted from each image slice, and hence there will be an additional summation over image slices. Moreover, in this case, $\hb{P}_j^H$ outputs a 2D signal of size $N \times N$. %from this vector.
%Finally, %$\hb{\tilde{F}}_{2D}\hb{P}^H\hb{Dz}$ is acquired by taking the
%2D Fourier transform of each slice of $\hb{P}^H\hb{\tilde{D}z}$ is taken to obtain the final form.
 
%When the image data is correlated along all directions, 
For the convolutional dictionary case, the last term $\hb{\tilde{F}}\hb{P}^H\hb{\tilde{D}z}$ can be efficiently computed using $\hb{P}=\hb{I}$ and the diagonalization property of the convolutional dictionaries. That is, the convolution matrix representing the $m$th dictionary filter can be expressed as $\hb{D}_m =  \hb{F}_{3D}^H \hb{\Theta}_m \hb{F}_{3D}$ where $\hb{F}_{3D}$ is the unitary 3D DFT matrix and $\hb{\Theta}_m$ is a diagonal matrix whose diagonal consists of the 3D DFT of the dictionary filter ${d}_m$ with $m = 1,\hdots,M$. As a result, the overall matrix $\hb{\tilde{D}} = [\hb{D}_1  \hdots \hb{D}_M ]$ can be written as $\hb{\tilde{D}} = \hb{F}_{3D}^H\hb{\Theta} \hb{\hat{F}}$  where $\hb{\hat{F}} = \hb{I}_M \Motimes \hb{F}_{3D}$ and $\hb{\Theta}$ is a matrix of $1 \times M$ blocks with each block given by $\hb{\Theta}_m$. By replacing this expression for $\hb{\tilde{D}}$ in the term $\hb{\tilde{F}}\hb{\tilde{D}z}$, the task becomes to compute $\hb{\tilde{F}}\hb{F}_{3D}^H\hb{\Theta} \hb{\hat{F}}\hb{z}$. Here multiplication by $\hb{\hat{F}}$ corresponds to taking the Fourier transforms of all 3D sparse codes for $m=1, \hdots, M$. % via the 3D FFT.
That is, $\hb{\hat{F}}\hb{z}=[ (\hb{F}_{3D}\hb{z}_1)^T | \hdots | (\hb{F}_{3D}\hb{z}_M)^T]^T$, where each term can be computed via the 3D FFT. Moreover, because $\hb{{\Theta}}$ is a block matrix consisting of diagonal matrices, multiplication by $\hb{{\Theta}}$ corresponds to element-wise 3D multiplications with the DFTs of the underlying dictionary filters and then summation.
%Then, for the multiplication $\hb{\Theta} \hb{\tilde{F}}_{3D}\hb{z}$, we simply take 3D transforms of dictionary filters $d_1[n_1,n_2,s],\hdots,d_M[n_1,n_2,s]$ and the sparse codes $z_1[n_1,n_2,s],\hdots,z_M[n_1,n_2,s]$, multiply them element-wise and then sum.
Lastly, the term $\hb{\tilde{F}}\hb{F}_{3D}^H$ can be simplified as
%Further simplification of % $\hb{\tilde{F}}_{2D}\hb{F}_{3D}^H$ is performed by decomposing $\hb{F}_{3D} = \hb{F}_D \Motimes \hb{F}_{2D}$ where $\hb{F}_D$ stands for one-dimensional DFT matrix. By using the properties of Kronecker product we obtain
$\hb{\tilde{F}}\hb{F}_{3D}^H = (\hb{I}_S \Motimes \hb{F}_{2D})(\hb{F}_{1D}^H \Motimes \hb{F}_{2D}^H) = \hb{F}_{1D}^H \Motimes \hb{I}_{N^2}$ using the properties of the Kronecker product, where $\hb{F}_{1D}$ stands for the 1D DFT matrix of size $S\times S$. %\be \label{kron}
%\begin{aligned}
%\hb{\tilde{F}}_{2D}\hb{F}_{3D}^H = (\hb{I}_S\Motimes \hb{F}_{2D})(\hb{F}_D^H \Motimes \hb{F}_{2D}^H) = \hb{F_D}^H \Motimes \hb{I},
%\end{aligned} 
%\ee
Hence multiplication by $\hb{\tilde{F}}\hb{F}_{3D}^H$ corresponds to %\hb{F}_D^H \Motimes \hb{I}_{N^2}$ 
taking 1D inverse DFTs along the third-dimension. %for each spatial index.

%For an image data with 2D correlations only, $\hb{D}_m = \hb{F}_{2D}^H \hb{\Theta}_m \hb{F}_{2D}$ and hence $\hb{\tilde{D}}$ can be written as $\hb{\tilde{D}} = \hb{\tilde{F}}^H \hb{\Theta} \hb{\bar{\tilde{F}}}$ where $\hb{\bar{\tilde{F}}}  = \hb{I}_{MS} \Motimes \hb{F}_{2D}$. Here $\hb{\Theta}$ is a matrix of $S \times SM$ blocks with each block is given by $\hb{\Theta}_m$ or $\hb{0}$ matrix. Also $\hb{\tilde{F}}\hb{P}^H\hb{\tilde{D}z} = \hb{\Theta} \hb{\bar{\tilde{F}}}\hb{z}$ where the multiplication by $\hb{\bar{\tilde{F}}}$ corresponds to taking the Fourier transforms of the sparse codes for $m = 1,\hdots,M$ and by $\hb{\Theta}$ corresponds to element-wise 2D multiplications and summation.
For the convolutional prior with 2D correlations only, computation of $\hb{\tilde{F}}\hb{\tilde{D}z}$ simplifies to taking the Fourier transforms of the 2D sparse codes $\hb{z}_{m,s}$ for $m = 1,\hdots,M$ and $s=1,\hdots,S$, computing element-wise 2D multiplications with the DFTs of the underlying dictionary filters and summing over $m$.

\subsubsection{Sparse code update}
%For minimization over $\hb{z}$, we need to solve the problem in Eq.~\eqref{iterd_ADMM2} 
The minimization in Eq.~\eqref{iterd_ADMM2} over the sparse code $\hb{z}$ 
%problem in Eq.~\eqref{iter_ADMM1} over the image $\hb{x}$ is 
has different solutions for patch-based and convolutional dictionaries. 

For the patch-based dictionary case, each sparse code $\hb{z}_j$ of the $j$th patch can be separately obtained as the solution of a least-squares problem: %Eq.~\eqref{iterd_ADMM2} is separable to least-squares problems which has the following closed-form solution: %whose solution is given by
\be \label{solz_patch2}
\begin{aligned}
    \hb{z}_j = (\rho I + \hb{{D}}^H\hb{{D}})^{-1}(\hb{{D}}^H\hb{P}_j\hb{x} + \rho(\hb{t}_j - \hb{u}_j)),
\end{aligned} 
\ee
where $\hb{t}_j$ and $\hb{u}_j$ are the auxiliary and dual variables corresponding to $\hb{z}_j$.

For the convolutional dictionary case, the resulting least-squares problem for $\hb{z}$ has the following normal equation:
\be \label{solz}
\begin{aligned}
    (\rho I + \hb{\tilde{D}}^H\hb{\tilde{D}})\hb{z} = \hb{\tilde{D}}^H\hb{x} + \rho(\hb{t} - \hb{u}).
\end{aligned} 
\ee
Note that when gradient regularization is used, a term that contains the convolution matrix $\hb{R}_i$, i.e. $\mu \sum_{i}\hb{{R}}_i^H\hb{{R}}_i$, should be added to the left-hand side. % of this equation. %with summation over $i$ is due to the gradient regularization. to the sparse codes, this step is slightly modified as
%Note that for an image data with 2D correlations only, $i = 1,2$.
%where $\hb{R}_i$ is convolution matrix of $\hb{r}_i$ for $i = 1,2,3$ can be diagonalized by the DFT matrices. Hence, the methods developed to solve Eq.~\eqref{solz} can be applied for solving Eq.~\eqref{solz_exten3D}. 
%%%REVISION: Burada Eq.~\eqref{solz} cozumunu anlatmak yerine, daha genel durum olan Eq.~\eqref{solz_exten3D} cozumunu anlatmak daha iyi olurdu. 
%%% Rlardan dolayı daha fazla variable kalabalıklığı olmasın diye bunu anlatmıştım. Aslında Riler de diagonal olduğu için \hb{I} yerine iki durum için de genellenebilir diagonal ve değişkene bağlı olmayan bir matris koysam sizin dediğiniz gibi olurdu. 

%Note that if we introduce gradient regularization to the sparse codes, this step is slightly modified as
%\be \label{solz_exten3D}
%\begin{aligned}
%    (\rho \hb{I} + \mu \sum_{i}\hb{{R}}_i^H\hb{{R}}_i + \hb{\tilde{D}}^H\hb{\tilde{D}})\hb{z} = (\hb{\tilde{D}}^H\hb{x} + \rho(\hb{t} - \hb{u})),
%\end{aligned} 
%\ee
Similar to the image update step, this normal equation can be solved efficiently through computations in the frequency domain.
By inserting the expression $\hb{\tilde{D}} = \hb{F}_{3D}^H\hb{\Theta} \hb{\hat{F}}$
where $\hb{\Theta}$ is a matrix of $1 \times M$ blocks with each block given by $\hb{\Theta}_m$ as before,
%used before, 
the following form can be obtained for efficient computation:
%the efficient computation of the sparse code update step:
%When the image data is correlated along all directions, the sparse code update in Eq.~\eqref{solz} is expressed in the frequency domain as
\noindent 
\be \label{solz_DFT}
\begin{aligned}
 (\rho I + \hb{{\Theta}}^H\hb{{\Theta}})\hb{\hat{F}}\hb{z} =\hb{{\Theta}}^H\hb{F}_{3D}\hb{x} + \rho \hb{\hat{F}}(\hb{t} - \hb{u}).
\end{aligned} 
\ee
Here the right-hand side can be computed as before via Fourier transforms, element-wise multiplications and summations. 
%%%REVISION: Burada acik acik tekrar soylenebilirdi ne islemler yapildigini. Bir önceki paragraflarda \hb{\hat{F}, \hb{F}_{3D}, \hb{{\Theta}} ile olan element-wise multiplicationları anlattığımız için okuyucu daha önceki açıklamalardan yola çıkarak bunu elde edebilir diye dusunduk. Sayfa sayısını azaltmaya çalıştigimiz icin yer vermedik.
For simplicity, let us denote the resulting vector from the right-hand side as $\hb{c}$, where $\hb{c} = [\hb{c}_{1}^T|\hdots| \hb{c}_{M}^T]^T$. Similarly, let us call $\hb{\hat{F}}\hb{z}=\hb{v}$, where $\hb{v} = [\hb{v}_{1}^T|\hdots| \hb{v}_{M}^T]^T$. Then Eq.~\eqref{solz_DFT} becomes $(\rho I + \hb{{\Theta}}^H\hb{{\Theta}}) \hb{v}=\hb{c}$. This linear system can be solved efficiently by replacing it with independent linear systems of size $M \times M$, each of which consists of a diagonal matrix plus a rank-one matrix~\cite{wohlberg2016}. 

Here the main idea is to swap the vector and entry indexing of the vectors $\hb{c}_m$ and $\hb{v}_m$, that is to convert $\hb{c}_m[n]$ to $\hb{\tilde{c}}_n[m]$ and $\hb{v}_m[n]$ to $\hb{\tilde{v}}_n[m]$. By applying the Sherman-Morrison formula, the solution is then given by~\cite{wohlberg2016}
\be \label{solz_DFT_SM}    
\begin{aligned}
\hb{\tilde{v}}_n = \rho^{-1}\left(\hb{\tilde{c}}_n - \frac{\hb{\tilde{\theta}}_{n}^H \hb{\tilde{c}}_n}{\rho + \hb{\tilde{\theta}}_{n}^H\hb{\tilde{\theta}}_{n} }\hb{\tilde{\theta}}_{n}\right),
\end{aligned} 
\ee
where $\hb{\tilde{\theta}}_{n}$ denotes the vector obtained by applying the same swapping operation on the diagonals of the matrices $\hb{\Theta}_m$'s. That is, if the vector $\hb{{\theta}}_{m}$ denotes the DFT of the $m$th dictionary filter, i.e. the diagonal of the matrix $\hb{\Theta}_m$, then swapping converts $\hb{{\theta}}_{m}[n]$ to $\hb{{\tilde\theta}}_{n}[m]$. Using Eq. \eqref{solz_DFT_SM}, $\hb{\tilde{v}}_n$'s can be obtained, and after rearranging, one can obtain $\hb{v}_m$'s, i.e. 3D DFTs of the sparse codes $\hb{z}_m$'s.

\subsubsection{Auxiliary variable update}
Similar to the analysis case, the minimization in Eq.~\eqref{iterd_ADMM3} over the auxiliary variable $\hb{t}$ is obtained through soft-thresholding:
\be
\label{softADMM}
\hb{t} = {\rm{soft}}(\hb{z} +\hb{u},\lambda/\rho).
\ee

\begin{algorithm}[t]
    \caption{Image reconstruction algorithm: synthesis case}
    \label{alg:algorithm_sum2}
        \textbf{Input:}  
        $\hb{y}$: measurement,
        $\hb{H}$: system matrix,
        $\hb{D}$: dictionary
        %$\hb{P}$: $\hb{I}$/patch-extractor
        \textbf{Output:}  
        $\hb{x}$: reconstructed image 
        \begin{algorithmic}[1]
        \State Choose $\lambda > 0$, $\rho > 0$, $\beta > 0$,  $\hb{t}^0$, and $\hb{u}^0$.
        \State  Compute $( t\hb{I} + \beta \hb{{\Lambda}}^H\hb{{\Lambda}})^{-1}$ and $\beta\hb{{\Lambda}}^H\hb{\bar{F}} \hb{y}$.
        \Repeat
        \State{update $\hb{x}^{l+1}$ using Eq.~\eqref{solx_synt_DFT}.}
        \State {update $\hb{z}^{l+1}$ using Eq.~\eqref{solz_patch2} or Eq.~\eqref{solz_DFT}.}
        \State{update $\hb{t}^{l+1}$ using Eq. \eqref{softADMM}.} 
        \State{update $\hb{u}^{l+1}$ as $\hb{u}^{l+1} = \hb{u}^l + \hb{z}^{l+1} - \hb{t}^{l+1}$.}
%        \If {$update$ $the$ $dictionary$}
%        \State{update $\hb{\tilde{D}}^{l+1}$, auxiliary and dual variables using Eq.~\eqref{ADMM0},~\eqref{ADMM1},~\eqref{subh} for patch-based or
%        Eq.~\eqref{d_admm},~\eqref{g_admm},~\eqref{h_admm} for convolutional dictionary.}
%        \Else
%        \State $\hb{\tilde{D}}^{l+1}$ = $\hb{\tilde{D}}^{l}$
%        \EndIf
        \State{$l \leftarrow l+1$}
        \Until{stopping criterion is satisfied.}
    \end{algorithmic}
\end{algorithm}

\subsubsection{Dictionary update} %Online dictionary update
\label{sec4}
This update has different forms for patch-based and convolutional dictionaries.

\paragraph{Patch-based dictionary update}
The minimization in Eq.~\eqref{patch_based_dic} over the dictionary $\hb{D}$ can be converted to an unconstrained problem by adding the normalization constraint to the objective function as a penalty: % function
%For dictionary-related updates, we first convert the problem in Eq.~\eqref{patch_based_dic} to an unconstrained problem by adding the constraint to the objective function as a penalty function
\be \label{Constrained1}
\begin{aligned}
    \underset{\hb{D}}{\min} \; ||\hb{D}\hb{Z}-\hb{X}||^2_F+\iota_{C_D}(\hb{D}),
\end{aligned} 
\ee
where $\hb{X} = [\hb{P}_1\hb{x}|\hdots| \hb{P}_J\hb{x}]$, $\hb{Z} = [\hb{z}_{1}|\hdots|\hb{z}_{J}]$, 
  and  $\iota_{C_D}(\hb{D})$ denotes the indicator function that takes value $0$ when the normalization constraint $||\hb{D}||_F = 1$ is satisfied, and $+\infty$ otherwise. After variable-splitting and expressing the problem in an augmented Lagrangian form, minimization over each variable is performed in an alternating fashion as follows: 
\be \label{ADMM0}
    \hb{D}^{l+1} =  \underset{\hb{D}}{\arg \min} \; \frac{1}{2} ||\hb{D}\hb{Z}^{l+1}-\hb{X}^{l+1}||^2_F+ \frac{\sigma}{2} ||\hb{D} - \hb{G}^{l} + \hb{E}^{l}||_F^2,
\ee
\be \label{ADMM1}
\begin{aligned}
\hb{G}^{l+1} =    \underset{\hb{G}}{\arg \min}\;  \iota_{C_D}(\hb{G}) + \frac{\sigma}{2} ||\hb{D}^{l+1} - \hb{G} + \hb{E}^{l}||_F^2,
\end{aligned} 
\ee
\be \label{subh}
\begin{aligned}
    \hb{E}^{l+1} = \hb{E}^l + \hb{G}^{l+1} - \hb{D}^{l+1},
\end{aligned} 
\ee
where $\hb{G}$ is the auxiliary variable and $\hb{E}$ is the dual variable in the ADMM framework.
The minimization in Eq.~\eqref{ADMM0} over the dictionary $\hb{D}$ corresponds to a least-squares problem with the
following closed-form solution:
% Here, the solution of this problem is obtained after taking the Hermitian of Eq.~\eqref{ADMM0} as
\be \label{sold_patch}
\begin{aligned}
    \hb{D} = (\hb{X}\hb{Z}^H + \sigma \hb{(G-E)}) (\hb{ZZ}^H + \sigma \hb{I})^{-1}.
\end{aligned} 
\ee
Lastly, the solution of %the problem in 
Eq.~\eqref{ADMM1} is obtained by geometry~\cite{afonso2010}: % as follows:
\begin{equation}
\hb{G} = \frac{(\hb{D}+ \hb{E})}{||{(\hb{D} + \hb{E})}||_F}.
\end{equation}
%Note that other dictionary learning algorithms such as K-SVD can also be exploited.
\paragraph{Convolutional dictionary update} The minimization in Eq.~\eqref{DL_Prior1} over the dictionary filters, $\hb{d}_m$, can be solved efficiently in the frequency domain. 
%For dictionary related updates, the problem in Eq.~\eqref{DL_Prior1} has to be solved. To efficiently solve this 
%in the frequency domain, 
For this, the following constraint set is defined for the filters: 
%in Eq.~\eqref{DL_Prior1} is modified as %follows:
\be \label{set}
\begin{aligned}
    C_d= \{\hb{d}_m \in \mathbb{R}^{N^2S} : (\hb{I} - \hb{QQ}^T)\hb{d}_m= 0,\quad ||\hb{d}_m||_2 = 1  \},
\end{aligned} 
\ee
with $\hb{Q}$ representing the zero-padding operator for %implicit 
%zero-padding of 
$\hb{d}_m$'s to the size of the sparse codes $\hb{z}_m$'s.  %required. % Explicit zero-padding 
%This operation is 
%mathematically expressed via multiplication by the zero-padding matrix $\hb{Q}$
Hence this set combines the normalization constraint with the spatial support constraint of the dictionary filters.
%For the image data with 3D correlations,
The problem in Eq.~\eqref{DL_Prior1} with this constraint set can then be converted to the following unconstrained problem:
%we concatenate the resulting matrices and vectors and convert the problem in Eq.~\eqref{DL_Prior1} to an unconstrained problem by adding the constraint to the objective function as a penalty function as %unconstrained version of Eq. ~\eqref{DL_Prior1} is expressed as
\be \label{Constrained2}
\begin{aligned}
    &\underset{\hb{d}}{\arg \min}\; ||\hb{Zd}-\hb{x}||^2_2 + \sum_{m} \iota_{C_d}(\hb{d}_m),
\end{aligned} 
\ee
where $\hb{d} = [\hb{d}_{1}^T|\hdots |\hb{d}_{M}^T ]^T$ is the vertically concatenated dictionary filter vector, and $\hb{Z} = [\hb{Z}_{1}| \hdots |\hb{Z}_{M} ]$ with $\hb{Z}_{m}$ denoting the convolution matrix for the sparse code $\hb{z}_{m}$. After variable-splitting and expressing the problem in an augmented Lagrangian form, minimization over each variable is performed in an alternating fashion as follows: %Moreover, $\hb{d}_m$ vectors are vertically concatenated as $\hb{d}$. 
%An auxiliary variable $\hb{g}$ and a dual variable $\hb{e}$ are inserted into Eq.~\eqref{Constrained2} to obtain the augmented Lagrangian form. Then, ADMM steps for this alternating minimization process are expressed as% ADMM steps of obtaining the solutions are expressed as
\be \label{d_admm}
\begin{aligned}
    \hb{d}^{l+1} = \underset{\hb{d}}{\text{argmin}} \;
    \frac{1}{2} ||\hb{Z{d}}-\hb{x}^{l+1}||^2_2+ \frac{\sigma}{2}||\hb{d}-\hb{g}^l + \hb{e}^l||^2_2,
\end{aligned} 
\ee
\be \label{g_admm}
%\begin{aligned}
    \hb{g}^{l+1} = \underset{\hb{g}}{\arg \min}\;\sum_{m}\hb{\iota}_{C_d}(\hb{g}_m) +
    \frac{\sigma}{2}||\hb{d}_m^{l+1}-\hb{g}_m + \hb{e}_m^{l+1}||^2_2,
%\end{aligned} 
\ee
\be \label{h_admm}
\begin{aligned}
    \hb{e}^{l+1} = \hb{e}^l + \hb{g}^{l+1} - \hb{d}^{l+1},
\end{aligned} 
\ee
where $\hb{g}_m$ and $\hb{e}_m$ are respectively the auxiliary and dual variables in the ADMM framework for $m = 1,\hdots,M$, $\hb{e}$ = $[\hb{e}_{1}^T|\hdots |\hb{e}_{M}^T ]^T$ and $\hb{g}$ = $[\hb{g}_{1}^T|\hdots |\hb{g}_{M}^T ]^T$. %The linear system providing the solution of least-squares problem for the dictionary update in Eq.~\eqref{d_admm} is

The minimization in Eq.~\eqref{d_admm} over the dictionary filter vector $\hb{d}$ corresponds to a least-squares problem with the following normal equation:
%For the solution of Eq.~\eqref{d_admm}, we face a least-squares problem which has the following normal equation
\be \label{sold_3D}
\begin{aligned}
    (\sigma\hb{I} + \hb{{Z}}^H\hb{{Z}})\hb{d} = \hb{{Z}}^H\hb{x} + \sigma(\hb{g} - \hb{e}).
\end{aligned} 
\ee
Here each convolution matrix $\hb{Z}_m$ %representing the $m$th dictionary filter 
can be decomposed as $\hb{Z}_m =  \hb{F}_{3D}^H \hb{\Gamma}_m \hb{F}_{3D}$ where $\hb{\Gamma}_m$ is a diagonal matrix whose diagonal consists of the 3D DFT of the sparse code $\hb{z}_m$. % with $m = 1,\hdots,M$. 
Hence the overall matrix $\hb{Z} = [\hb{Z}_{1}| \hdots |\hb{Z}_{M} ]$ can be expressed as $\hb{{Z}} = \hb{F}_{3D}^H\hb{{\Gamma}}\hb{\hat{F}}$ where $\hb{\Gamma}$ is a matrix of $1 \times M$ blocks with each block given by $\hb{\Gamma}_m$. Following the same steps with the solution of Eq. \eqref{solz},
%$\hb{{Z}}$ = $\hb{F}_{3D}^H\hb{{\Gamma}}\hb{\hat{F}}$ %to Eq.~\eqref{sold_3D}
%\be \label{sold_DFT_3D}
%\begin{aligned}
 %   (\sigma\hb{I} + \hb{{\Gamma}}^H\hb{{\Gamma}})\hb{\tilde{F}}_{3D}\hb{d} = \hb{{\Gamma}}^H\hb{F}_{3D}\hb{x} + \sigma\hb{\tilde{F}}_{3D}(\hb{g} - \hb{e}),
%\end{aligned} 
%\ee
%The left-hand side consists of the sum of a block diagonal rank-$1$ matrix and a diagonal matrix,
this equation is solved in a similar way in the frequency domain via Sherman-Morrison formula. 
%%%REVISION: Burada bir cumleyle nasil bir hesap yapildigi aciklanmaliydi, butunluk acisindan. Didem, hepsini söyledigimizi dusunuyor.
Note that for an image data with 2D correlations only, the problem in Eq.~\eqref{Constrained2} %is formulated as
will be changed to include an
additional summation over the image slice index $s$. In this case, the resulting minimization over $\hb{d}$ %similar to the problem in Eq.~\eqref{d_admm} is not solvable
cannot be solved via efficient Sherman-Morrison formula. Instead, iterated Sherman-Morrison formula, conjugate-gradient method, spatial tiling or consensus framework can be used~\cite{wohlberg2016, cardona2018}. In this work, we use iterated Sherman Morrison formula for this purpose.
%to perform minimization over $\hb{d}$~\cite{wohlberg2016}.
%\be \label{Constrained2_2d}
%\begin{aligned}
%    &\underset{\hb{d}_{m}}{\arg \min}\; \sum_{s=1}^{S}||\hb{Z}_s\hb{d}-\hb{x}_s||^2_2 + \sum_{m} \iota_{C_d}(\hb{d}_m),
%\end{aligned} 
%\ee

Lastly, the solution of Eq.~\eqref{g_admm} is obtained by geometry as
\begin{equation}
\hb{g}_m = \frac{\hb{QQ}^T(\hb{d}_m + \hb{e}_m)}{||\hb{QQ}^T(\hb{d}_m + \hb{e}_m)||_2}.
\end{equation}
Here $\hb{Q}^T$ operation crops an $L \times L \times R$ (or $L \times L$) data and $\hb{Q}$ operation zero-pads this cropped data to the size $N \times N \times S$ (or $N \times N$) when the convolutional prior is used for 3D (or 2D) correlations.
%there is correlation all through
%the image data (or only in 2D). 
%Buradaki Q matrisi, onceden bahsi gecen Q matrisi, yani zero-padding operatorünü temsil eden matris.

\subsection{ADMM parameter update}
For the selection of the penalty parameter $\rho$, the following adaptive strategy is employed~\cite{boyd2011}:
\be
\label{paramS}
\rho^{l+1} =
\begin{cases}
    \tau \rho^l \ \text{if}\ ||r^{l}||_2 >\mu||s^{l}||_2,\\
    \rho^l / \tau \ \text{if}\ ||s^{l}||_2 >\mu||r^{l}||_2,\\
    \rho^l \ \text{otherwise,}
\end{cases}
\ee
where $s^{l} = \rho^l ||t^l - t^{l-1}||_2$ and $r^{l} = ||z^{l} - t^{l}||_2$ are primal and dual residuals, respectively, and the parameters are chosen as $\tau = 2$ and $\mu = 10$.
The same strategy is also used to update the parameter $\sigma$ in the dictionary update. Moreover, the stopping criterion is chosen as $||x^{l+1} - x^l||_2/||x^l||_2 < 10^{-4}$.

\section{Numerical Results}
We now present numerical simulations to illustrate the performance of the developed reconstruction algorithms with different priors and compare with each other. % .
To illustrate their performance, these algorithms are applied to three-dimensional reconstruction problems in computational spectral imaging, and their performance is numerically demonstrated for various cases with or without correlation along the third dimension. % Introda kullanılan bir cümle, önceki cümleyle benzeşiyor biraz
\subsection{Case with no correlation along the third dimension}  
%2D Case
%Case with 2D correlations only
%Case with no correlation along the third dimension
The performance is first illustrated %for a problem 
in photon sieve spectral imaging (PSSI)~\cite{oktem2014,kamaci2017} for a %an image 
multi-spectral data with 2D
spatial correlations.
For this, we consider a spectral dataset of size $128 \times 128 \times 3$ ($3$ EUV wavelengths between $33.3-33.5$ nm with $0.1$ nm interval)
%Goruntu boyutu icin 128 cok kucuk aslinda, ama artik boyle biraktik.
constructed %using solar images 
from NASA's database of solar images~\cite{nasaWebsite}. %%% at 335 nm 
%%%REVISION: Burada 335nm deki goruntuleri alip, bu 3 farkli dalgaboyunun goruntuleri gibi kullaniyoruz. Bunu belirtmek belki iyi olabilirdi.

\newcommand{\widthFig}{0.5in}
\newcommand{\widthFiga}{1.5in}
\begin{figure*}[t]
    \begin{center}    \label{fig:allIm}

        \subfloat[]{\includegraphics[width=\widthFig]{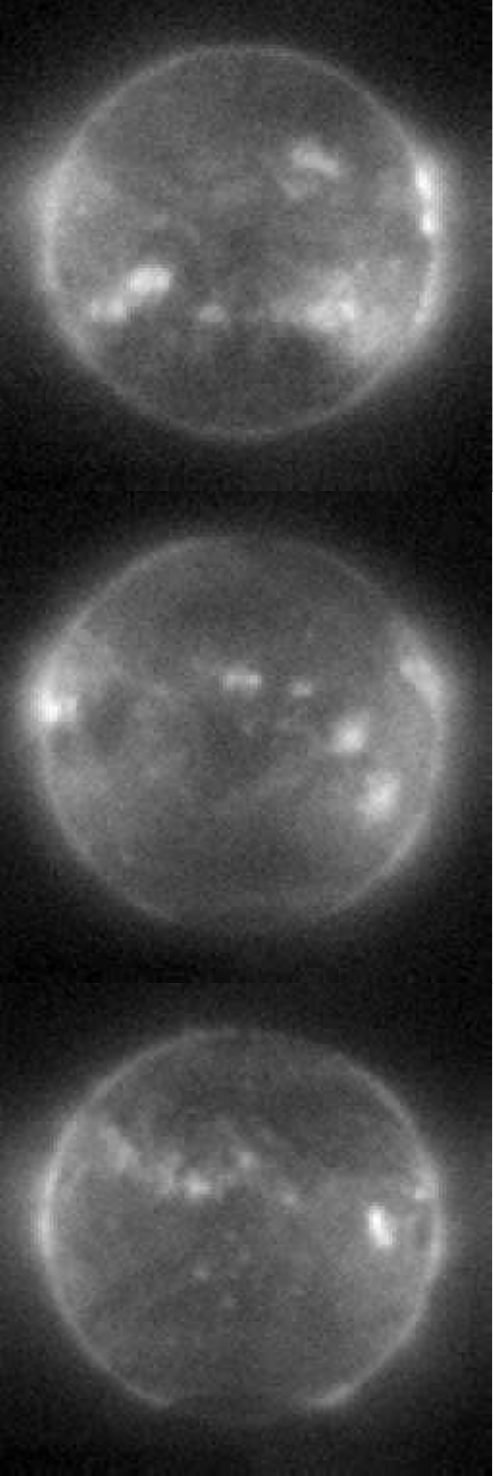}\label{fig:measured}}
        \hspace{0.005in}
        \subfloat[]{\includegraphics[width=\widthFiga]{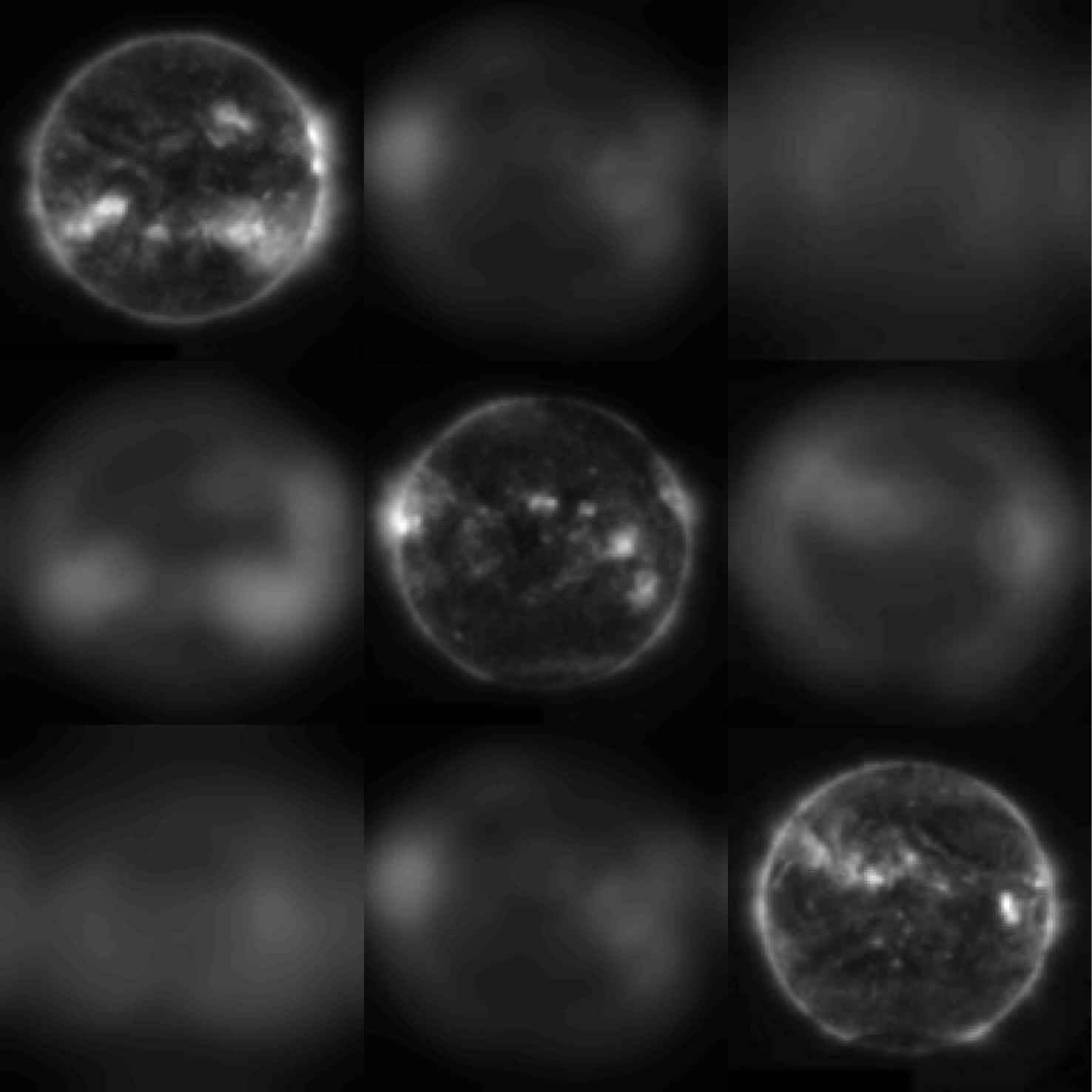}\label{fig:cont}}
        \hspace{0.005in}
        \subfloat[]{\includegraphics[width=\widthFiga]{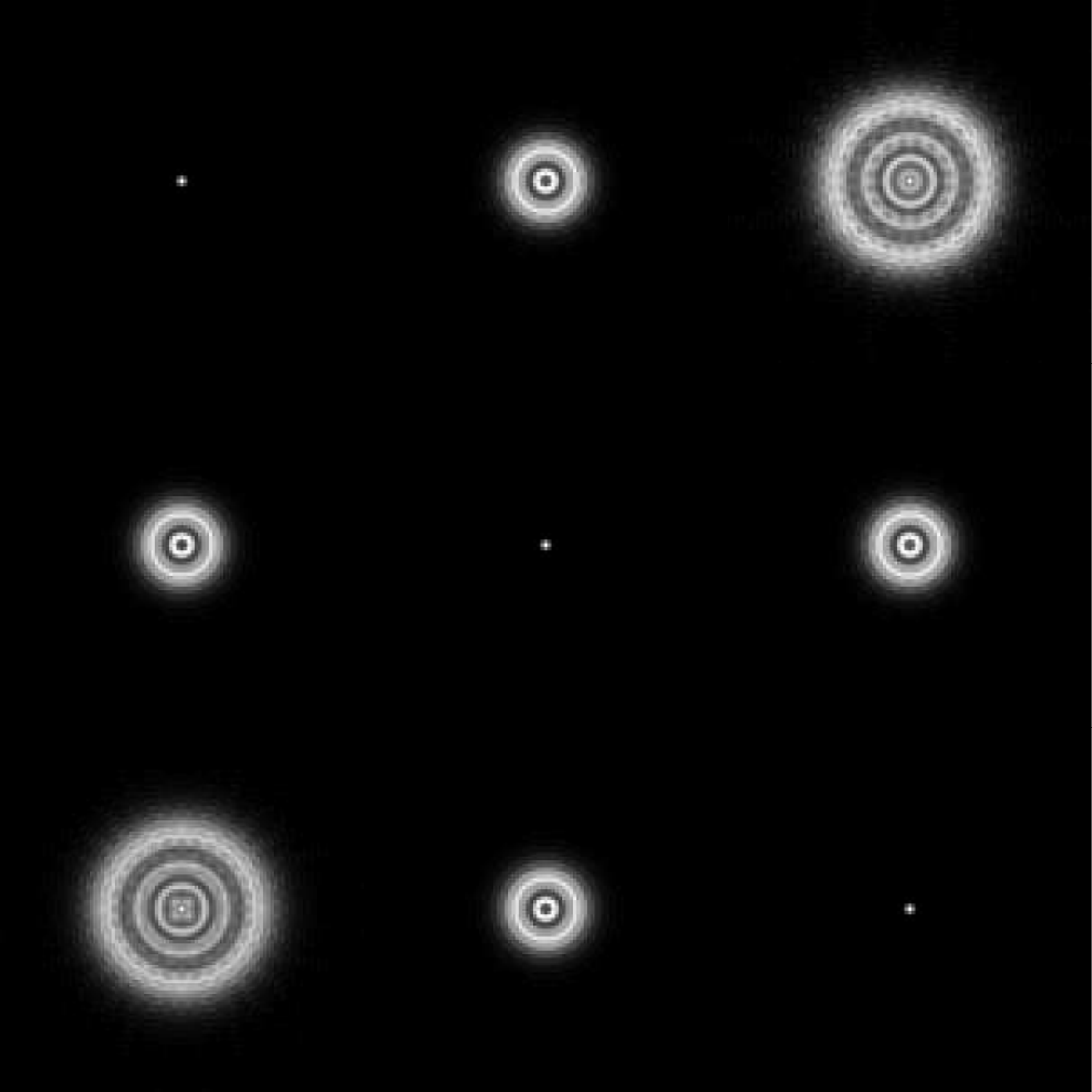}\label{fig:psf}}
        \hspace{0.005in}    
        \subfloat[]{\includegraphics[width=\widthFig]{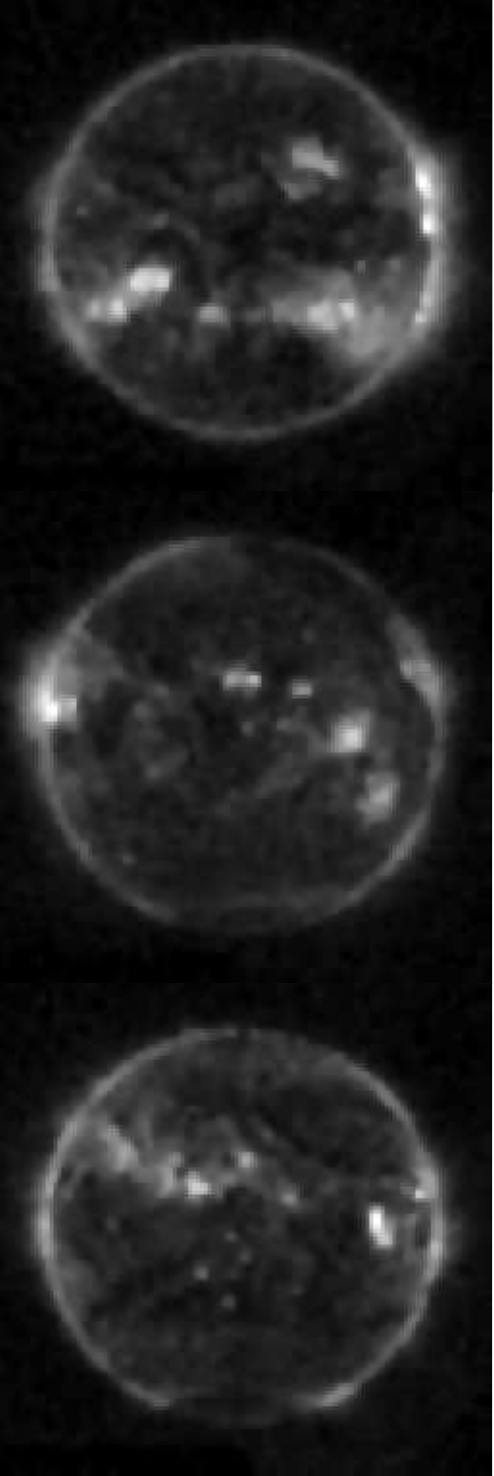}\label{fig:reco1}}
        \hspace{0.005in}
        \subfloat[]{\includegraphics[width=\widthFig]{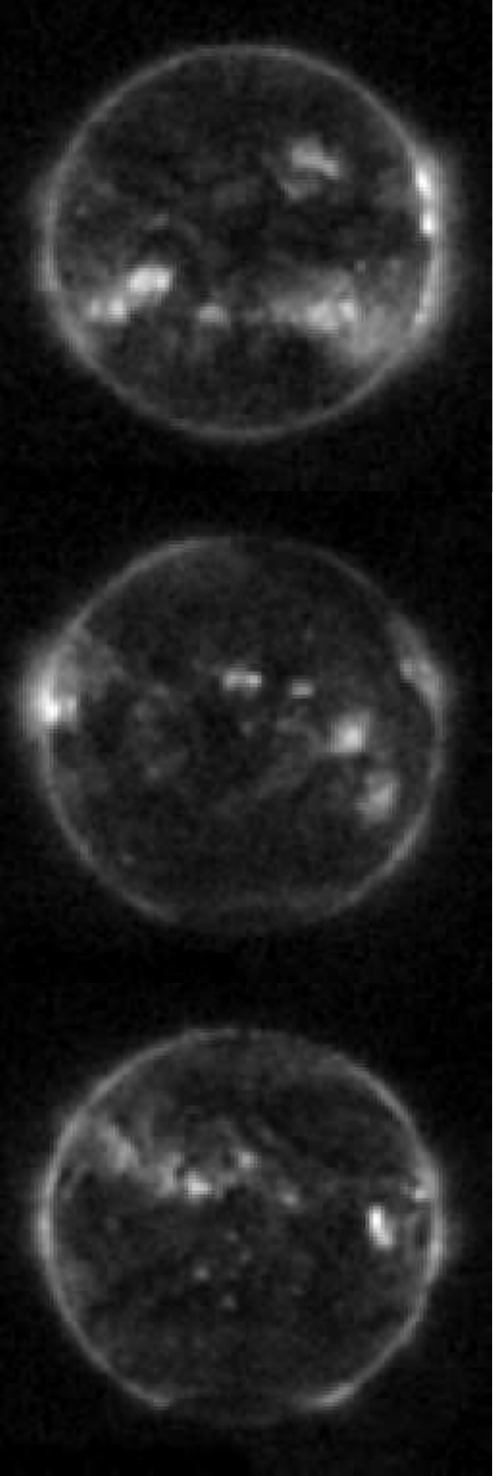}\label{fig:reco2}}
        \hspace{0.005in}
        \subfloat[]{\includegraphics[width=\widthFig]{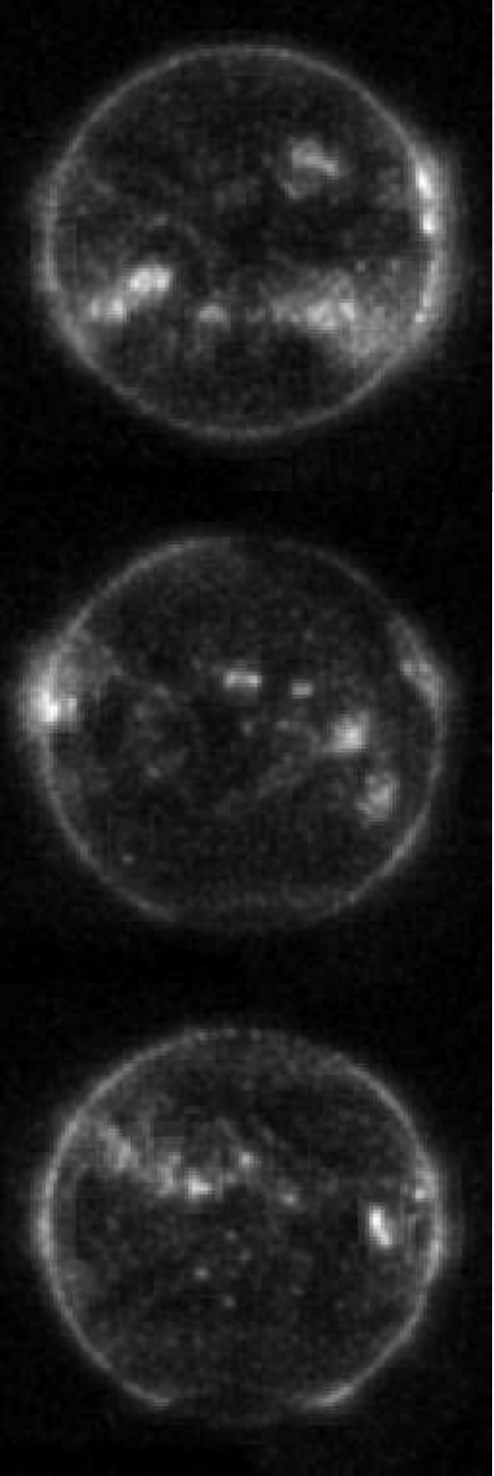}\label{fig:reco3}}
        \hspace{0.005in}
         \subfloat[]{\includegraphics[width=\widthFig]{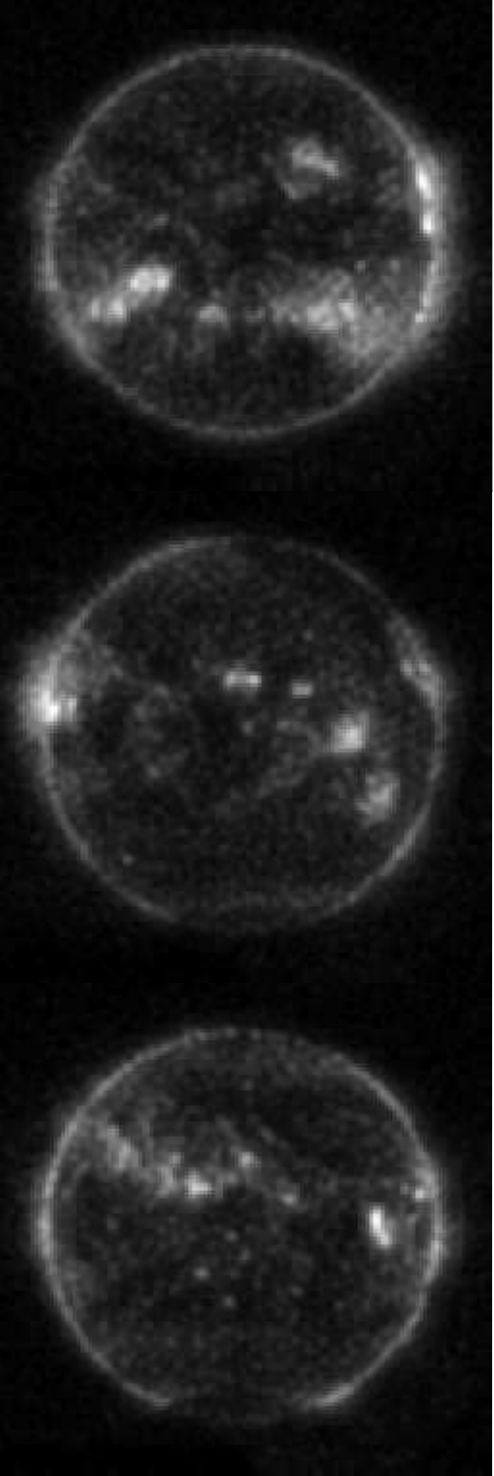}\label{fig:reco4}}
        \hspace{0.005in}
        \subfloat[]{\includegraphics[width=\widthFig]{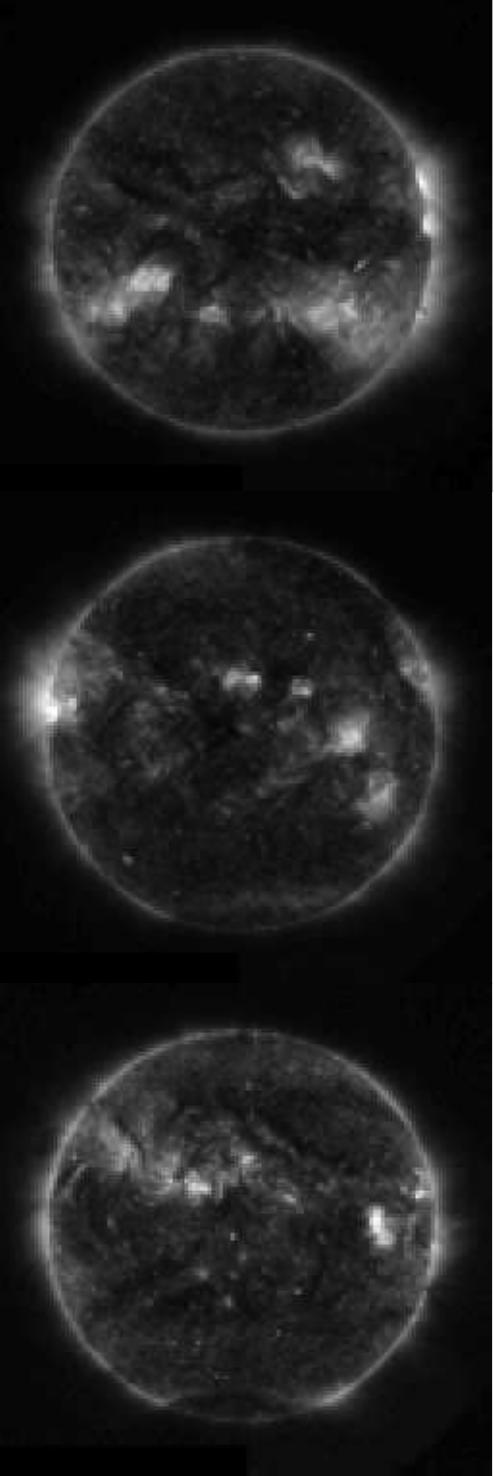}\label{fig:orig}}
        \hspace{0.005in}    
        \caption{(a) Measured images for SNR = $20$ dB, (b) Contributions of each spectral component to the measurements, (c) Acting PSFs on spectral images, (d) Reconstructed images with TV, (e) Online patch-based dictionary, (f) Online convolutional dictionary, (g) Online convolutional dictionary with Tikhonov regularization and (h) Original Images.}
    \end{center}
\end{figure*}

\begin{table}[b]
    %\label{ParSelect}
    \caption{Parameters used for different priors.}
    \centering
    \resizebox {4.05in}{!}{
        \begin{tabular}{|c|c|c|c|c|c|c|c|c|c|}
            \hline 
            
            Parameter & \multicolumn{3}{c|}{TV}& \multicolumn{3}{c|}{PatchDic }& \multicolumn{3}{c|}{ConvDic }\\             \hline
            SNR (dB)&$20$&$30$&$40$&$20$&$30$&$40$&$20$&$30$&$40$ \\             \hline
            
            $\lambda$& $10^{-2}$ & $10^{-3}$ & $10^{-4}$&
            \multicolumn{3}{c|}{$0.05$}
            &$0.2$&$0.15$&$0.15$ \\ \hline

            $\beta$& \multicolumn{3}{c|}{$1$} &$100$ & $500$&$3000$&$2$&$8$&$50$  
            \\ \hline
            $\rho$ & \multicolumn{3}{c|}{$10\lambda$ }  &\multicolumn{3}{c|}{$1$ }& \multicolumn{3}{c|}{$50\lambda+0.5$}\\ \hline
            $\sigma$ & \multicolumn{3}{c|}{$-$ } & \multicolumn{3}{c|}{$1$ } &  \multicolumn{3}{c|}{$10$ }\\ \hline
            $\mu$ & \multicolumn{3}{c|}{$-$ } & \multicolumn{3}{c|}{$-$ } &  \multicolumn{3}{c|}{$0.01$ (Tikhonov) }\\ \hline
        \end{tabular}    
    }
    \label{table:par}
\end{table}

\begin{table*}[t!]
    \caption{Comparison of reconstruction PSNRs (dB) and SSIMs for different priors and SNRs.}
       \centering
    {      \resizebox {6in}{!}{
   
            \begin{tabular}{|c|c|c|c|c|c|c|c|}
                \hline
                &  & %Wavelet  & PatchDic & 
                PatchDic& PatchDic & ConvDic  & ConvDic & ConvDic& ConvDic\\ 
                SNR (dB) & TV  &%Transform& Inverse &
                KSVD &Online Update & Offline &  Offline & Online Update  & Online Update \\
                
                & %&& DCT
                &&& & Tikhonov &  &   Tikhonov\\ \hline
%                &&&&&&&&& \\
%                $15$ &$30.57/0.80$ & $28.22/0.68$ &$30.21/0.81$&$30.76/0.81$ &$31.33/0.80$ & $30.33/0.80$ &$30.68/0.82$ &$30.55/0.81$&$30.95/0.83$\\ \hline   
                %&&&&&&&&& \\ 
                $20$ &$32.62/ {0.88}$ & % $31.34/0.85$ &$31.80/0.85$&
                $32.45/0.86$ & $ {32.97}/0.86$& $32.17/0.85$
                &$32.18/0.86$
                &$32.57/0.87$ &$32.67/0.87$ \\ \hline
%                &&&&&&&&& \\
%                $25$ &$34.22/0.91$ &$34.47/0.86$ &$33.61/0.88$&$33.81/0.88$ & $34.56/0.89$ &$34.13/0.89$&$34.50/0.91$ &$34.36/0.90$&$34.51/0.91$\\ \hline
            %    &&&&&&&&& \\
                $30$ & $36.25/0.94$ 
                %$35.74/0.94$ 
                & %$33.69/0.88$&$35.12/0.91$&
                $35.42/0.91$& $36.23/0.93$ & $36.00/0.93$ &
                $36.43/0.93$ &
                $36.42/0.94$ &  ${36.52/0.94}$ \\ \hline
%                &&&&&&&&& \\
%                $35$ & $37.58/0.96$ & $34.95/0.91$ &$36.38/0.93$&$36.68/0.94$ &$36.67/0.94$ & $37.57/0.95$  &$37.88/0.96$ &$37.80/0.95$&$38.01/0.96$ \\ \hline
                %&&&&&&&&& \\
                $40$ &$39.57/0.97$ & %$36.37/0.93$&$38.09/0.95$&
                $38.42/0.95$ &$39.21/0.96$ & $39.21/0.96$
                &$39.63/0.97$
                &$39.80/0.97$  &$ {39.83/0.97}$\\ \hline
            \end{tabular}
    }}
    \label{table:PSNR1}
\end{table*}

For the photon sieve, a sample design~\cite{davila2011} for EUV solar imaging is considered, with the smallest hole diameter of $5$ $\mu$m and the outer diameter of $25$ mm. Photon sieve is a diffractive lens whose focal length changes with the incoming wavelength. 
%photon sieve has wavelength-dependent focal length.
%%%REVISION: Burada PSSI sistemi icin bir gorsel vermek faydali olurdu.
The PSSI system takes measurements at the focal planes of each of these three wavelengths,
%The photon sieve system records the intensities at the three focal planes corresponding to each wavelength, 
that is at
%This results in a photon sieve with first-order focal lengths of 
$f_1$ = $3.754$ mm, $f_2$ = $3.742$ mm, and $f_3$ = $3.731$ mm. Then
at the first focal plane, $f_1$, the measurement contains the focused
image of the first spectral component at wavelength $\lambda_1=33.3$ nm, overlapped with the defocused spectral images of the remaining two %spectral 
components (at wavelengths $\lambda_2=33.4$ nm and $\lambda_3=33.5$ nm). %and vice versa at the other focal planes.
Pixel size of the detector is chosen as \SI{2.5}{\micro\metre} to match the diffraction-limited %spatial 
resolution of the imaging system. 

The measurements are simulated using the forward model in Eq. \eqref{forwardModel} with white Gaussian noise. % where $K = S = 3$. 
Here the number of measurements and the number of unknown spectral images are $K = S = 3$. 
%After the field passes through the photon sieve, we capture the measurements at the focal planes of each wavelengths.
Fig.~\ref{fig:measured} shows the resulting measurements at the three focal planes %together 
with the contributions of each spectral component shown in Fig.~\ref{fig:cont}. These contributions are obtained by convolving the original spectral images in Fig. \ref{fig:orig}, with the corresponding PSFs in Fig. \ref{fig:cont}.
%Sample measurements at the three focal planes are shown in Fig. \ref{fig:measured}, together with the contributions from each spectral image to these measurements shown in \ref{fig:cont}. 
The acting PSFs for the three spectral components are computed using the available PSF formula for the photon sieve~\cite{oktem2014, oktem2018}.
%%%REVISION: Burada ICIP2013'e de referans verilebilirdi.
These PSFs illustrate the different amount of blur acting on spectral components. Hence the measurements involve not only the superposition of all spectral components but also significant amount of blur. %and degradation.

%The photon sieve system takes measurements at the focal planes of each of these wavelengths,$128 \times 128$ solar EUV images shown in Fig. \ref{fig:orig} are used as the true spectral images at these wavelengths, and the measurements are generated using the forward model in Eq.~\eqref{forwardModel} with signal-to-noise ratios (SNRs) of $20$, $30$ and $40$ dB. Each measurement is the superposition of differently blurred spectral images. Here Fig. \ref{fig:measured} represents measured intensities with 20 dB SNR. The contributions from each spectral image to these measurements are shown in Fig. \ref{fig:cont}. These contributions are acquired by convolving point spread functions in Fig. \ref{fig:psf} with original images in Fig. \ref{fig:orig}. 

To analyze the performance with different noise levels, SNRs of $20$, $30$ and $40$ dB are considered. Reconstructions are obtained from the noisy measurements using  Algorithm \ref{alg:algorithm_sum1} and \ref{alg:algorithm_sum2} with 2D priors. %As a prior information,
The parameters used for different priors are listed in Table~\ref{table:par}.
For the analysis case, we exploit 2D isotropic TV, %and Wavelet  
which takes $20$ seconds for image reconstruction on a computer with 8 GB of RAM and i7 7500U 2.70 GHz CPU. 

Secondly, we exploit a %pre-trained
patch-based dictionary (PatchDic) %learned 
%for the patch-based dictionary (PatchDic)
with no online dictionary update. This dictionary is trained offline using the K-SVD algorithm~\cite{aharon2006} with $16$ representative solar images taken from the same database. We also use %include 
a randomly initialized patch-based dictionary, which is updated online throughout the iterations.  %, which is adaptively learned from the data. %The synthesis model requires size optimization of the dictionaries. %The dictionary-based image reconstruction requires the size optimization.  
For both %with and without online dictionary update, 
cases, the number of patches extracted from each image is $N^2$ = $16384$ with one-stride. 
The patch size is numerically optimized as $6 \times 6$, based on the simulations performed for $20$ dB SNR, which results in a dictionary size of $36 \times 36$. Image reconstruction %with $K = P = 3$ 
takes around $150$ and $200$ seconds with offline and online updated dictionary, respectively. %($150$ seconds without dictionary update). 

Moreover, we utilize convolutional dictionaries (ConvDic) in a similar manner. When an online dictionary update is not performed, the convolutional dictionary is trained offline with the same $16$ solar images.
%using the algorithm presented in Section \ref{convalgorithm} by removing the image update steps. 
The %convolutional 
dictionary size is numerically optimized as $12 \times 12$ and the number of filters as %set to
$M = 4$. Using this prior, a single reconstruction takes approximately $10$ and $20$ seconds with offline and online updated dictionary, respectively. % ($10$ seconds without dictionary update), 
%which is substantially faster than the patch-based one. % On the other hand, when online dictionary update is performed, convolutional dictionary is randomly initialized. 
The same experiments are also repeated by adding the gradient (Tikhonov) regularization, which result in similar reconstruction time.

\begin{table*}[t!]
    \caption{Comparison of reconstruction PSNRs (dB) / SSIMs / SAMs for different priors, datasets and SNRs.}
    \centering
    {
        \resizebox {6in}{!}{
            \begin{tabular}{|c|c|c|c|c|c|}
                \hline
                
                Dataset &SNR (dB) & 2D Wavelet $\Motimes $ 1D DCT& PatchDic& ConvDic& ConvDic (Tikhonov)\\\hline
                
                \multirow{3}{*}{Objects}    
                &$20$
                &$24.82/11.78^{\circ}/0.82$ & $25.42/11.13^{\circ}/0.77$ %$25.05/ 11.34^{\circ}/0.83$ %$24.88/ 11.79^{\circ}/0.81$
                &$26.71/10.16^{\circ}/0.77$ & $ {26.82/9.64^{\circ}/0.84}$\\ 
                &$30$&$26.18/10.57^{\circ}/0.88$ & $26.78/10.03^{\circ}/0.85$ &$28.42/8.64^{\circ}/0.85$ &$ {28.54/8.49^{\circ}/0.89}$\\ 
                &$40$&$27.76/9.30^{\circ}/0.89$  & $28.91/8.35^{\circ}/0.87$ &
                $29.63/7.71^{\circ}/0.89$&
                $ {30.15/7.44^{\circ}/0.91}$ \\ \hline
                
%                \multirow{3}{*}{Beads}
%                &$20$ &$23.16/13.81^{\circ}/0.79$ & $25.45/13.73^{\circ}/0.83$&$24.94/11.96^{\circ}/0.84$ \\
%                &$30$&$25.66/12.04^{\circ}/0.88$  &$28.31/11.67^{\circ}/0.88$ &$28.10/9.96^{\circ}/0.90$\\ 
%                &$40$&$28.11/11.08^{\circ}/0.93$ & $29.83/10.45^{\circ}/0.93$ &$29.46/9.03^{\circ}/0.94$\\ \hline
                
                \multirow{3}{*}{Flowers}
                &$20$&$27.94/20.06^{\circ}/0.77$& $28.82/21.61^{\circ}/0.80$
                & $28.16/21.26^{\circ}/0.78$&$ {28.31/19.76^{\circ}/0.79}$\\ 
                &$30$&$31.09/ {16.05^{\circ}}/0.87$& $31.57/17.68^{\circ}/0.88$ &
                $30.47/18.64^{\circ}/0.85$ &$ {31.19}/16.86^{\circ}/ {0.88}$ \\
                &$40$&$33.46/15.04^{\circ}/0.92$  & $ {35.16/13.26^{\circ}/0.93}$ &$33.12/14.94^{\circ}/0.92$ &$ {34.09/13.83^{\circ}/0.93}$  \\ \hline
                
                \multirow{3}{*}{Pompoms}    
                &$20$&$28.02/ {9.40^{\circ}}/0.83$ & $28.66/10.27^{\circ}/0.82$
                &$ {28.54}/11.49^{\circ}/0.81$&$28.48/10.43^{\circ}/ {0.83}$\\
                &$30$&$29.29/ {9.01^{\circ}}/0.87$ & $ {30.44}/9.20^{\circ}/0.88$ 
                & $29.57/10.80^{\circ}/0.86$  &$ {30.21}/9.41^{\circ}/ {0.88}$\\ 
                &$40$&$30.62/8.85^{\circ}/0.91$ & $ {32.33}/8.15^{\circ}/ {0.92}$
                &$30.40/10.13^{\circ}/ 0.90$&$ {31.41/8.71^{\circ}/ 0.92}$\\ \hline
                
                \multirow{3}{*}{Threads}
                &$20$&$28.83/11.78^{\circ}/0.84$&$28.86/12.11^{\circ}/0.85$
                &$29.51/12.46^{\circ}/0.81$ &$29.39/11.82^{\circ}/0.85$\\
                &$30$&$31.14/10.79^{\circ}/0.90$&$32.52/10.44^{\circ}/0.90$ 
                &$31.82/11.79^{\circ}/0.87$&$32.27/10.29^{\circ}/0.90$ \\ 
                &$40$&$34.04/9.32^{\circ}/0.94$&$34.87/9.34^{\circ}/0.95$%$31.82/14.67^{\circ}/0.84$
                &$34.12/9.72^{\circ}/0.93$ &$34.29/9.17^{\circ}/0.95$\\ \hline
        \end{tabular}}
    }
    \label{table:PSNR3D_overall}    
\end{table*}
%To reconstruct the images, analytical sparsifying transforms, patch-based dictionaries, and convolutional dictionaries with and without Tikhonov regularization are exploited. 

The average reconstruction performance for all cases is given in Table \ref{table:PSNR1} in terms of PSNR and SSIM.
%In Table \ref{table:PSNR1}, we present the image reconstruction PSNR and SSIM values for different SNRs, i.e., $20$ dB to $40$ dB for comparison of different analysis and synthesis models. 
%These results are obtained as taking the average of 
These average values are computed through $10$ Monte-Carlo runs for $4$ different %solar image sets. 
spectral (solar) data sets. %The optimal parameter range for the algorithms are given in Table \ref{table:par}. 
As seen from the table, PSNR is always above $32$ dB, and SSIM is above $0.85$, which demonstrate faithful reconstructions for all cases. 
%Here, the reconstruction PSNRs are above $30$ dB even when SNR is $20$ dB and increases significantly with increasing SNR. 
Moreover, randomly initialized dictionary is effectively adapted to the data through online dictionary update, and yields higher PSNR and SSIM than the offline case for both patch-based and convolutional dictionaries.

To also visually evaluate the results, we provide sample reconstructions for SNR=$20$ dB case in Fig. \ref{fig:reco1}, \ref{fig:reco2}, \ref{fig:reco3} and \ref{fig:reco4}, %the reconstructed spectral images 
together with the true images in Fig. \ref{fig:orig}. 
%For SNR=$20$ dB, the highest performance is attained via a patch-based dictionary with online dictionary update for PSNR and TV regularization for SSIM. However, convolutional dictionary with Tikhonov regularization and online dictionary update has slightly better performance than the other alternatives for less noisy regimes. 
Although TV prior, patch-based and convolutional dictionaries with online dictionary update provide similar reconstruction performance with comparable PSNR and SSIM values, visual comparison suggests that the image details are better preserved in the convolutional dictionary case.
%As seen, the image details and edges, as well as the spectral variations, are well preserved in the reconstructions. Note that the convolutional dictionary (regular and Tikhonov) recovers the characteristic features of each spectral images better than the other counterparts.
Moreover, convolutional dictionary and TV prior result in similar reconstruction times, whereas patch-based dictionary is approximately $10 \times$ slower.
\subsection{Case with correlations in three dimensions} % all directions
%all through the image data
The performance is now illustrated in PSSI system for a spectral data with 3D  correlations. For this, we consider spectral datasets of size $256 \times 256 \times 16$ ($16$ wavelengths between $510-660$ nm with $10$ nm interval)
taken
from online spectral database referred as Objects~\cite{nascimento2002}, Flowers~\cite{yasuma2010}, Pompoms~\cite{yasuma2010} and Threads~\cite{yasuma2010}. For the photon sieve design, the smallest hole diameter is chosen as \SI{15}{\micro\metre} and the outer diameter is $3.51$ mm. Moreover, the pixel size of the detector is chosen as \SI{7.5}{\micro\metre} \ to match the diffraction-limited %spatial 
resolution of the imaging system. As before, the PSSI system takes measurements at the focal planes of each of these sixteen wavelengths. For example, for the wavelength at $580$ nm
the focal length is $9.08$ cm. As a result, each measurement contains the focused image of one of the spectral components, %at wavelength $\lambda_1=580$ nm, 
overlapped with the defocused spectral images of the remaining fifteen %spectral 
components. 
\begin{figure}[b!]
    \begin{center} 
        \includegraphics[width=1.55in]{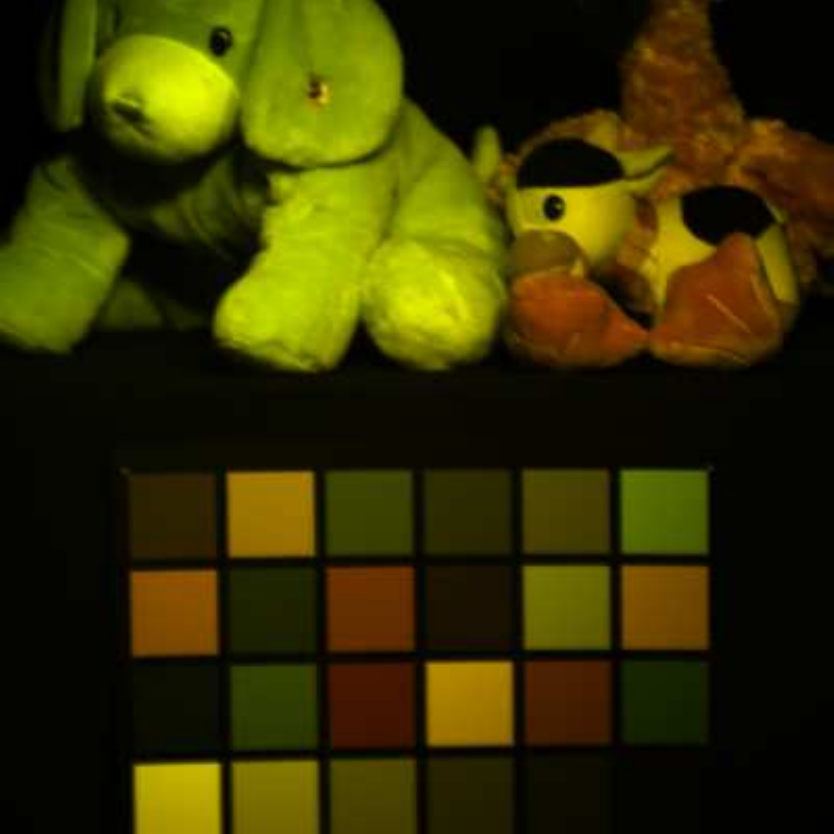}
        \caption{Datacube used for training.} \label{fig:toys}
    \end{center}
\end{figure}
%Bu goruntuler Oguzhan'in makalesindeki Fig.3 gibi gozukmeli. Oguzhan'in cizim icin kullandigi kod: 
% hcolored_script_for_superimposed_31bands.m

The measurements are simulated again using the forward model in Eq. \eqref{forwardModel} with white Gaussian noise. % where $K = S = 3$. 
Here the number of measurements and the number of unknown spectral images are $K = S = 16$. To analyze the performance with different noise levels, SNRs of $20$, $30$ and $40$ dB are considered as before. Reconstructions are obtained from these noisy measurements using %the algorithms in 
Algorithm \ref{alg:algorithm_sum1} and \ref{alg:algorithm_sum2} with 3D priors. The parameters used for different priors are listed in Table~\ref{table:par2}.

\begin{table}[b]
    %\label{ParSelect}
    \caption{Parameters used for different priors.}
    \centering
    \resizebox {3.3in}{!}{
        \begin{tabular}{|c|c|c|c|c|c|c|c|c|c|}
            \hline 
            
            Parameter & \multicolumn{3}{c|}{Transform}& \multicolumn{3}{c|}{PatchDic }& \multicolumn{3}{c|}{ConvDic }\\             \hline
            SNR (dB)&$20$&$30$&$40$&$20$&$30$&$40$&$20$&$30$&$40$ \\             \hline
            
            $\lambda$& $0.5$ & $0.1$ & $0.01$&
            \multicolumn{3}{c|}{$0.0001$}
            &\multicolumn{3}{c|}{$0.001$}\\ \hline
            
            $\beta$& \multicolumn{3}{c|}{$1$} &$0.1$ & $1$&$10$&$0.01$&$0.1$&$0.2$  
            \\ \hline
            $\rho$ & \multicolumn{3}{c|}{$500\lambda$ }  &\multicolumn{3}{c|}{$1000$ }& \multicolumn{3}{c|}{$1000$}\\ \hline
            $\sigma$ & \multicolumn{3}{c|}{$-$ } & \multicolumn{3}{c|}{$10$ } &  \multicolumn{3}{c|}{$10$ }\\ \hline
            $\mu$ & \multicolumn{3}{c|}{$-$ } & \multicolumn{3}{c|}{$-$ } &  \multicolumn{3}{c|}{$0.1$ (Tikhonov) }\\ \hline
        \end{tabular}    
    }
    \label{table:par2}
\end{table}

Similar to the earlier spectral imaging approaches \cite{willett2014,kar2019}, for the analysis case, we exploit a Kronecker basis as $\hb{T}$ = $\hb{T}_1 \Motimes \hb{T}_2$ where $\hb{T}_1$ is the basis for 2D Symmlet-$8$ wavelet and $\hb{T}_2$ is the 1D discrete cosine (DCT) basis. This transformation is computed by first taking the %2D Symmlet-$8$ 
wavelet transform of each spectral image and then 1D DCT along the spectral dimension. In this case,  image reconstruction %with $K = P = 3$ 
takes around 17 minutes. %$1000$ seconds. 
%Note that the parameters in Eq.~\eqref{analysis_ADMM} are adjusted as $\beta$ = $1$,  $\rho = 500 \lambda$ and $\lambda$ = $\{0.5, 0.1, 0.01\}$ for $\{20, 30, 40\}$ dB SNR.% for $20$ dB, $\lambda$ = $0.1$ for $30$ dB and $\lambda$ = $0.01$ for $40$ dB. %Due to substantial cost of patch-based dictionaries, we did not include them in this large-scale inverse problem. 

%We first exploit a 3D transform for the image reconstructions: Kronecker basis $\hb{T}$ = $\hb{T}_1 \Motimes \hb{T}_2$ where $\hb{T}_1$ is the basis for 2D Symmlet-$8$ wavelet and $\hb{T}_2$ is the 1D cosine basis. 

 % For the photon sieve, a sample design used with the outer diameter of the photon sieve as $3.51$ nm and the diameter of the smallest hole as $15$ $\mu$m, resulting in a focal length of $9$ cm at $560$ nm.
%We take measurements at the focal planes corresponding to these wavelengths.  
%The photon sieve system takes measurements at the focal planes of each of these wavelengths. The number of measurements $K = 16$ and different SNR levels are considered. The measurements are generated using the forward model in Eq.~\eqref{forwardModel} with signal-to-noise ratios (SNRs) of $20$, $30$ and $40$ dBs.
\begin{figure*}[t!]
    \begin{center} 
        \subfloat[]{\includegraphics[width=\textwidth]{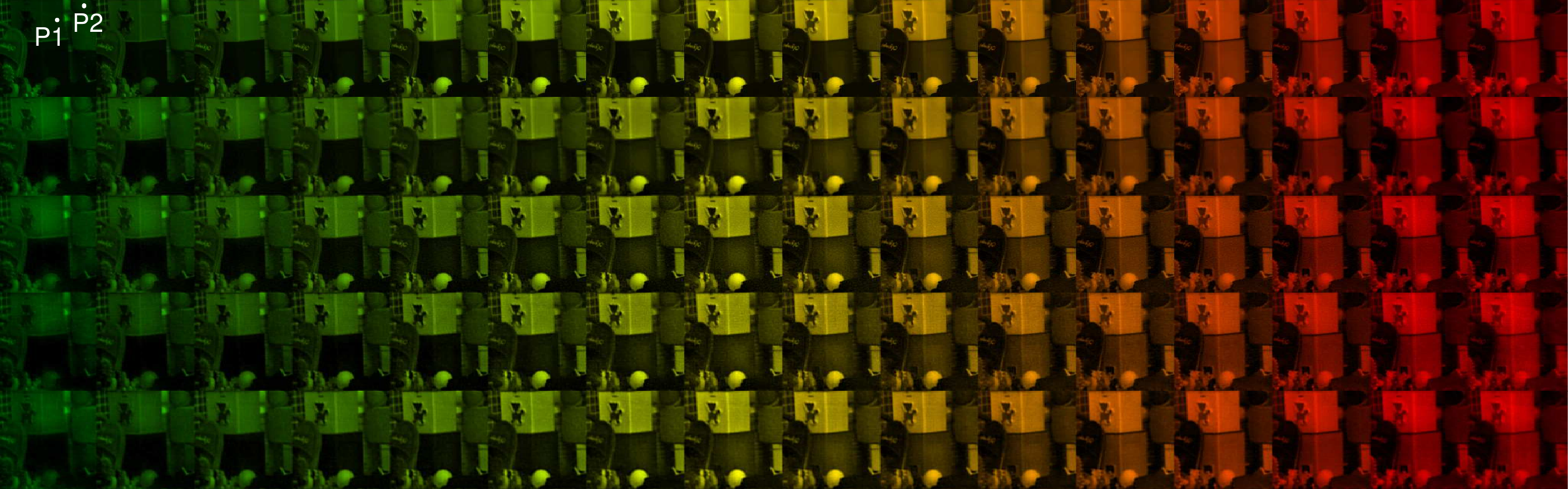}\label{fig:objects}}
        \hfil 
        \subfloat[]{\includegraphics[width=\textwidth]{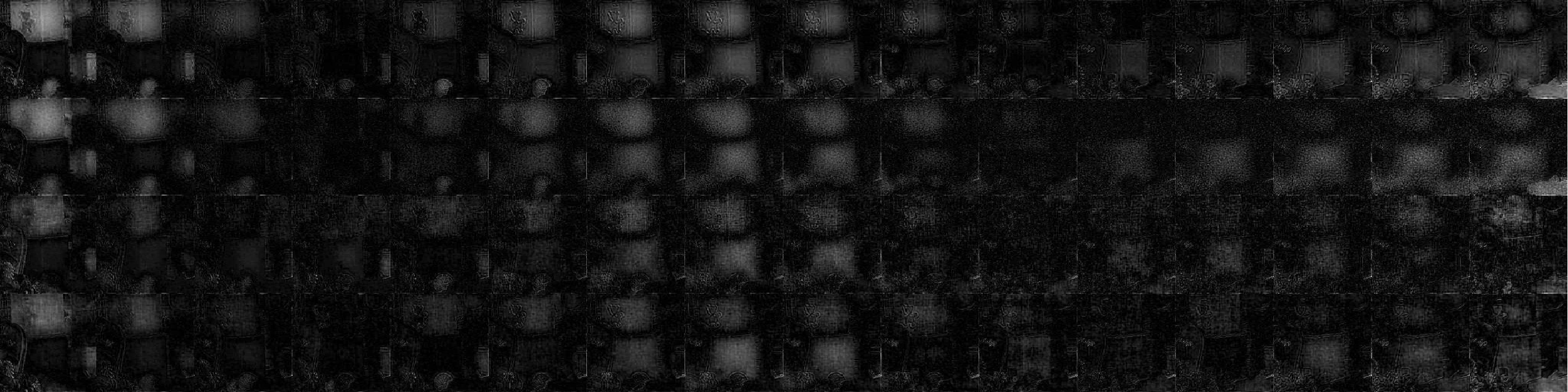}\label{fig:objects_diff}}
        \hfil
        \caption{(a) Top-to-bottom: Original Images, reconstructed spectral images using 2D Symmlet $\Motimes $ 1D DCT transform, patch-based dictionary, convolutional dictionary and convolutional dictionary with Tikhonov regularization for SNR = $20$ dB, (b) The difference between original image and reconstructed images. %
        }        
    \end{center}
\end{figure*}

Secondly, we exploit a patch-based dictionary with online dictionary update. The initial dictionary is trained offline using the K-SVD algorithm with $25$ spectral datacubes of size $256 \times 256 \times 16$ cropped from Toys data~\cite{yasuma2010} shown in Fig. \ref{fig:toys}. Here the patch size is chosen as $6 \times 6 \times 16$, resulting in a dictionary of 
size $576 \times 576$. The number of patches extracted from each datacube is $N^2$ = $65536$ %since we extracted the patches 
with one-stride only in spatial dimensions. In this case, image reconstruction %with $K = P = 3$ 
takes around $100$ minutes. %$5800$ seconds. %($150$ seconds without dictionary update). 
%The parameters in Eq.~\eqref{synthesis_ADMM} and \eqref{ADMM0} are adjusted as $\rho$ = $1000$, $\sigma$ = $10$,  $\lambda$ = $0.0001$ and $\beta$ = $\{0.1, 1, 10\}$ for $\{20, 30, 40\}$ dB SNR. %$\beta$ = $0.01$ for $20$ dB, $\beta$ = $0.1$ for $30$ dB, and $\beta$ = $0.2$ for $40$ dB. 
%Here,  $\rho$ and  $\sigma$ are iteratively updated using Eq.~\eqref{paramS}.

Lastly, we utilize convolutional dictionaries. %in a similar manner. 
The initial convolutional dictionary is trained offline with the same $25$ spectral datacubes and then an online dictionary update is performed %, which is updated online 
throughout the iterations.
%using the algorithm presented in Section \ref{convalgorithm} by removing the image update steps. 
The %convolutional 
dictionary size is numerically optimized as $32 \times 32 \times 5$ and the number of filters as %set to
$M = 6$. Using this prior, a single reconstruction takes approximately $35$ minutes. %$2100$ seconds.  The same experiments are also repeated by adding the gradient (Tikhonov) regularization, which result in similar reconstruction time. Here PSNRs, SAMs, and SSIMs are given in Table \ref{table:comp3D} and \ref{table:comp3D2} with respect to changing number of dictionary filters (NOF) and dictionary size with Tikhonov regularization for SNR=$20$ dB.
%Furthermore, the parameters
%in Eq.~\eqref{synthesis_ADMM} and \eqref{d_admm} are adjusted as $\rho$ = $1000$, $\sigma$ = $10$,  $\lambda$ = $0.001$ and $\beta$ = $\{0.01, 0.1, 0.2\}$ for $\{20, 30, 40\}$ dB SNR. %$\beta$ = $0.01$ for $20$ dB, $\beta$ = $0.1$ for $30$ dB, and $\beta$ = $0.2$ for $40$ dB. 
%Here,  $\rho$ and  $\sigma$ are iteratively updated using Eq.~\eqref{paramS}. Also, $\mu$ = $0.1$ with Tikhonov regularization.

The average reconstruction performance for all cases is given in Table \ref{table:PSNR3D_overall} in terms of PSNR, SSIM, and spectral angular mapper (SAM)~\cite{park2007}.
%%%REVISION: Bu tanimlarin tamamini (SNR, PSNR, SAM,...) makalenin icinde verelim.
These average values are computed through $10$ Monte-Carlo runs for each dataset. As seen from the table, the performance of different priors varies for different datasets and SNRs. In fact, each prior provides different capabilities over spatial and spectral dimensions.

\begin{figure}[b!]
    \begin{center} 
        \subfloat[]{\includegraphics[width=0.5\textwidth]{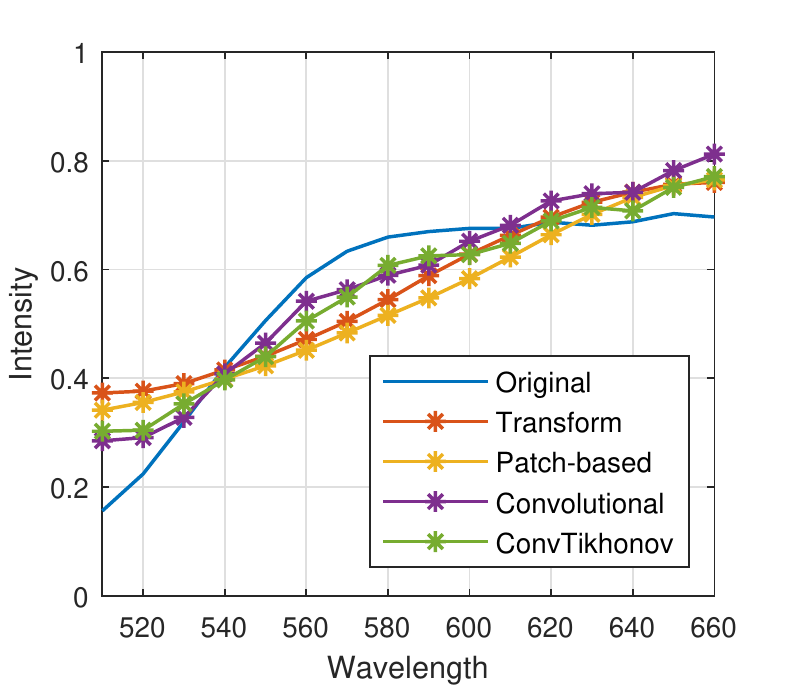}\label{fig:P1}}
        \hfil
        \subfloat[]{\includegraphics[width=0.5\textwidth]{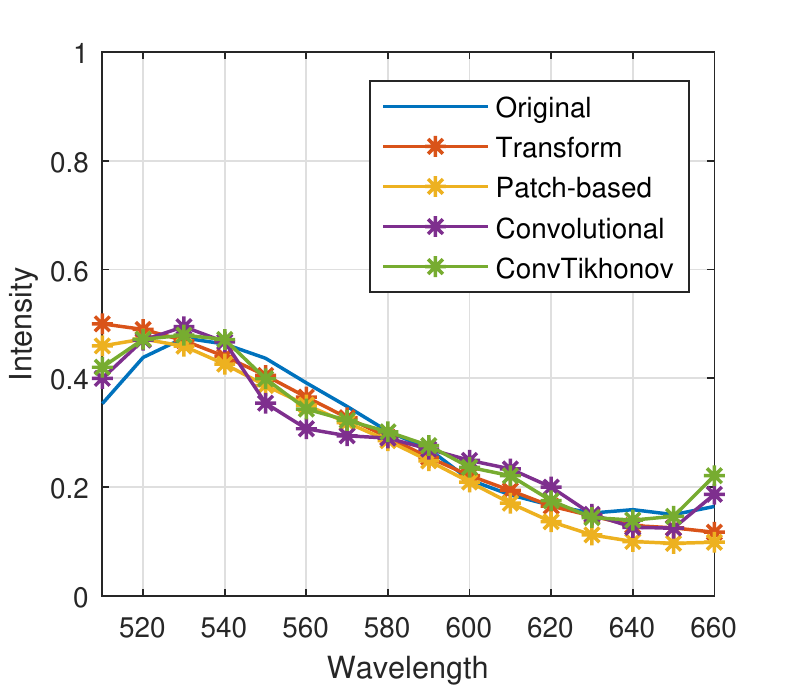}\label{fig:P2}}
        \hfil
        \caption{(a) Spectra at the points P1, and (b) P2 for SNR = $20$ dB.}
    \end{center} %
\end{figure}

To visually evaluate the results, we provide sample reconstructions in Fig. \ref{fig:objects} for the Objects dataset and SNR = $20$ dB case, together with the true images.
%The reconstructed Objects data using 2D Symmlet $\Motimes $ 1D DCT transform, patch-based dictionary, convolutional dictionary, and convolutional dictionary with Tikhonov regularization are shown in Fig. \ref{fig:objects} for the sixteen sources, together with the original scenes for comparison, when the input SNR is $20$ dB.
For easier interpretation and comparison of results, the absolute difference between the original spectral images and the reconstructed ones are shown in Fig.~\ref{fig:objects_diff} as well.
To investigate the %successful 
recovery along the spectral dimension, we also select two representative points with different spectral characteristics, as shown as P1 and P2 in Fig. \ref{fig:objects}. The reconstructed spectra of these points are plotted in Fig. \ref{fig:P1} and \ref{fig:P2}, together with the original spectra. 

These results demonstrate that the chosen transform prior (2D Symmlet $\Motimes $ 1D DCT) generally yields smoother reconstructions over space and spectrum. Hence it can work fine if the original image data has smooth variations; however, this is generally not the case. As a result, this analysis prior often causes the largest errors due to the loss of image details along spatial and spectral directions, which can also be observed from higher SAM or lower PSNR/SSIM values.  

On the other hand, with the patch-based dictionary, the spatial details are generally preserved better, but now there is additional unwanted grainy structure in space. Nevertheless, it often achieves the highest PSNR and SSIM. However, same is not true for the spectral recovery. The spectra recovered with the patch-based dictionary are generally overly smooth, resulting in the worst reconstruction performance along the spectral dimension and the highest SAM values. One possible cause here is the chosen patch size, which does not perform partitioning in the spectral dimension. But note that the reconstruction with this patch size is already $3 \times$ slower than the convolutional dictionary and $6 \times$ slower than the transform-based alternative. Hence working with smaller patches along spectrum will bring much higher computational cost. 
%%%REVISION: Aslinda 1-stride yerine daha buyuk bir stride kullanilsa ve spectrum yonunde de birkac patch tanimlansa, reconstruction performansinin iyilesmesi ve reconstruction time'in ayni kalmasi ihtimali var bence. Hatta space yonunde 6 yerine daha buyuk boyutta patch almak da daha iyi calisabilir gibi gozukuyor bana. Ama yorumlarimi mecburen mevcut sonuclar uzerinden yapiyorum. Emin olamadigim icin, bu ihtimallerden de simdilik bahsetmiyorum. Ama hakemler gundeme getirebilir. Ayrica convolutional icin boyutu 32x32x5 olarak secip, patch-based icin 6x6x16 ile calismak da garip gozukuyor, justification'i zor. Neden 6x6x16 olarak secildigiyle ilgili bir aciklama da vermiyoruz zaten. Bu asamada artik buna da takilmiyorum ama hakemler buyuk bir ihtimalle takilacaktir.

The results suggest that convolutional dictionary provides a better trade-off between reconstruction performance and time compared to the patch-based one. With the %used
convolutional prior, the resulting errors are more uniform over space and spectrum. That is, both spatial and spectral characteristics (variations) are generally well-preserved in the reconstructions, as can also be seen from high PSNR/SSIM and low SAM values. As expected, the inclusion of Tikhonov regularization yields a smoother reconstruction and less grainy spatial structure, but may come with the slight cost of loss of some spatial details.

\section{Conclusion}
In this paper, we have developed a unified framework for the solution of a general class of inverse problems, namely convolutional inverse problems,
that are widely encountered in multidimensional imaging. 
Considering a general image-formation model and using ADMM, we developed fast image reconstruction algorithms 
%with analysis and synthesis sparsity models.
that can exploit different analysis and synthesis priors as well as correlations in different dimensions.
%that can exploit sparse models in analysis or synthesis forms. 
In the analysis case, 
%discrete derivative operators or sparsifying transforms have been utilized.
multidimensional sparsifying operators are utilized. In the synthesis case, convolutional or patch-based dictionaries %have been utilized, 
are exploited and adapted to correlations in different dimensions.
%Based on the available prior knowledge about the image of interest, there may be correlations either all through the image data %in all directions or only in certain dimensions.
%The inverse problem has been formulated and the resulting optimization problems have been solved via the alternating direction method of multipliers (ADMM). 
% The obtained reconstruction algorithms have efficient and closed-form update steps.

To illustrate their utility and versatility, the developed algorithms with different priors are applied to 3D reconstruction problems in computational spectral imaging, and their performance is comparatively evaluated for various cases with and without correlation along the third dimension. 
Although analysis priors performed best in terms of reconstruction time, 
%convolutional dictionary has an advantage over the patch-based one but is inferior to the sparsifying-transform.
image details were generally better %reconstructed 
preserved with dictionary-based priors.
The results suggest that convolutional dictionary provides a better trade-off between reconstruction performance and time.
%the performance of the algorithms are very similar and varies with respect to the reconstructed image data.
Future work can focus on exploiting other structured dictionaries such as those with
tensor or Kronecker structure~\cite{caiafa2013,semerci2014,soltani2016,xie2018}. 
% dantas2017,friedland2014,qi2016,

%To illustrate their performance, these algorithms have been applied to three-dimensional reconstruction problems in computational spectral imaging, and their performance is numerically demonstrated for various cases with or without correlation along the third dimension. For an image data with two dimensional correlations only, we have shown that the convolutional dictionary with Tikhonov regularization has slightly better performance than the isotropic TV and patch-based dictionary. For an image data with three dimensional correlations, the results suggest that the performance of the algorithms are very similar and varies with respect to the reconstructed image data. We also observe that PSNR, SAM, and SSIM performances are enhanced with Tikhonov regularization for the convolutional dictionary. Considering the reconstruction time, convolutional dictionary has an advantage over the patch-based one but is inferior to the sparsifying-transform.

%The versatile ADMM-based reconstruction algorithms developed in this paper are broadly applicable to various different imaging modalities involving convolutional inverse problems.
The versatile ADMM-based reconstruction algorithms developed in this paper are broadly applicable to linear shift-variant imaging systems whose response slowly varies across the field of view, time, depth, or spectral dimensions. % as widely encountered in multidimensional imaging. 
%image reconstruction problems in multidimensional imaging
Moreover, the algorithms can be parallelized and easily extended to use with other priors such as those based on deep learning. %~\cite{}, 
%which can enable even faster reconstruction. % with good performance. Future work will focus on this.
As the advent of multidimensional imaging modalities expands to perform sophisticated tasks, these algorithms are essential for fast iterative reconstruction in various large-scale problems.
%To conclude, this paper opens up new possibilities for fast iterative reconstruction in large-scale imaging problems.
%for large-scale inverse problems  in various imaging problems.  
%for large-scale image reconstruction 

%As future work, the performance of these priors can be evaluated for the compressive setting, where there are fewer measurements than the unknowns. Furthermore, for comparison, deep learning priors can be exploited for convolutional inverse problems.

%%%REVISION: Asagidaki bilgi ilk sayfada veriliyor. Asagidaki section'a gerek olup olmadigini kontrol edelim.
\section*{Acknowledgement}
This work is supported by the Scientific and Technological Research Council of Turkey (TUBITAK) under grant 117E160 (3501 Research Program).

%\clearpage

\bibliographystyle{unsrt}  

\bibliography{main} 

\end{document}